\documentclass[aps,prappl,twocolumn,groupedaddress,showkeys,showpacs,superscriptaddress,floatfix]{revtex4-1}
\usepackage{epsfig}
\usepackage{multirow}
\usepackage{amsmath, amssymb,mathtools}
\usepackage{color}
\usepackage{graphicx}
\usepackage{dcolumn}
\usepackage{bm}
\usepackage{subfigure}
\usepackage[utf8]{inputenc}
\usepackage[T1]{fontenc}

\begin{document}

\title{Hysteretic superconducting heat-flux quantum modulator}

\author{Claudio Guarcello\thanks{e-mail: claudio.guarcello@nano.cnr.it}}
\affiliation{SPIN-CNR, Via Dodecaneso 33, 16146 Genova, Italy}
\affiliation{NEST, Istituto Nanoscienze-CNR and Scuola Normale Superiore, Piazza S. Silvestro 12, I-56127 Pisa, Italy}
\affiliation{Radiophysics Department, Lobachevsky State University, Gagarin Ave. 23, 603950 Nizhni Novgorod, Russia}
\author{Paolo Solinas\thanks{e-mail: paolo.solinas@spin.cnr.it }}
\affiliation{SPIN-CNR, Via Dodecaneso 33, 16146 Genova, Italy}
\author{Massimiliano Di Ventra\thanks{e-mail:}}
\affiliation{Department of Physics, University of California, San Diego, La Jolla, California 92093, USA}
\author{Francesco Giazotto\thanks{e-mail: giazotto@sns.it}}
\affiliation{NEST, Istituto Nanoscienze-CNR and Scuola Normale Superiore, Piazza S. Silvestro 12, I-56127 Pisa, Italy}

\date{\today}

\begin{abstract}
We discuss heat transport in a thermally-biased SQUID in the presence of an external magnetic flux, when a non-negligible inductance of the SQUID ring is taken into account. A properly sweeping driving flux causes the thermal current to modulate and behave hysteretically. The response of this device is analysed as a function of both the hysteresis parameter and degree of asymmetry of the SQUID, highlighting the parameter range over which hysteretic behavior is observable. Markedly, also the temperature of the SQUID shows hysteretic evolution, with sharp transitions characterized by temperature jumps up to, e.g., $\sim0.02\text{K}$ for a realistic Al-based setup. In view of these results, the proposed device can effectively find application as a temperature-based superconducting memory element, working even at GHz frequencies by suitably choosing the superconductor on which the device is based.
\end{abstract}


\maketitle

\section{Introduction}
\label{Intro}\vskip-0.2cm

In 1965, Maki and Griffin~\cite{Mak65} predicted that in a temperature-biased Josephson junction (JJ) the flow of an electronic heat current should depend on the macroscopic phase difference between the superconductors forming the junction. Recently, the features of the phase-coherent thermal transport in Josephson devices have been investigated and confirmed experimentally in several interferometer-like structures~\cite{Gia12,Mar13,MarGia13,Mar14,MarSol14,For16,ForTim16,ForGia16}. In Ref.~\cite{Gia12}, for instance, the thermal counterpart of a symmetric DC SQUID with negligible inductance of the loop was demonstrated. The heat current flowing through a thermally biased SQUID depends on the external magnetic flux $\Phi$, i.e., $\propto \left | \cos(\pi\Phi/\Phi_0) \right |$. In Ref.~\cite{Gia12}, clear modulations of the drain temperature as a function of $\Phi$ were observed, due to the interference between the coherent components of the heat currents flowing throughout the JJs forming the SQUID. The coherent nature of the thermal current was further confirmed in Ref.~\cite{Mar14} by the observation of thermal diffraction patterns in a flux driven, temperature-biased ``short'' rectangular tunnel JJ. When a temperature-biased extended JJ is threaded by a magnetic flux, a Fraunhofer-like diffraction pattern, i.e., $\propto \left | \sin(\pi\Phi/\Phi_0)/(\pi\Phi/\Phi_0) \right |$, for the drain temperature is observed~\cite{Mar14}.

In this paper, we theoretically investigate the thermal transport in a temperature-biased SQUID with a non-negligible ring inductance, as a slowly changing external magnetic flux (i.e., in the adiabatic regime) is taken into account. We show that hysteresis in the thermal current comes to light for proper values of the system parameters. By considering a simple thermal model accounting for the thermal currents flowing in/from the cold electrode of the SQUID~\cite{GiaMar12}, the modulation, due to the external flux, of its temperature, in both nonhysteretic --i.e., with vanishing inductance-- and hysteretic regimes, is explored. Notably, we predict that also the temperature behaves hysteretically, showing sudden transitions as the number of enclosed flux quanta in the SQUID ring changes. When this occurs, clear thermal jumps, up to $\Delta T_2\sim0.02\text{K}$, are observed. 

The paper is organized as follows. In Sec.~\ref{ModelResults} the theoretical background used to describe a thermally-biased SQUID is presented. The thermal currents are introduced and studied by varying the values of appropriate parameters, for several temperatures of the SQUID branches. 
In Sec.~\ref{ThermalModel}, the behavior of the temperature of the cold electrode of the SQUID is explored, as the thermal contact with bath phonons is taken into account. In Sec.~\ref{Conclusions} conclusions are drawn.

\section{Model and Results}
\label{ModelResults}\vskip-0.2cm 
\begin{figure*}[htpb!!]
\centering
\includegraphics[width=\textwidth]{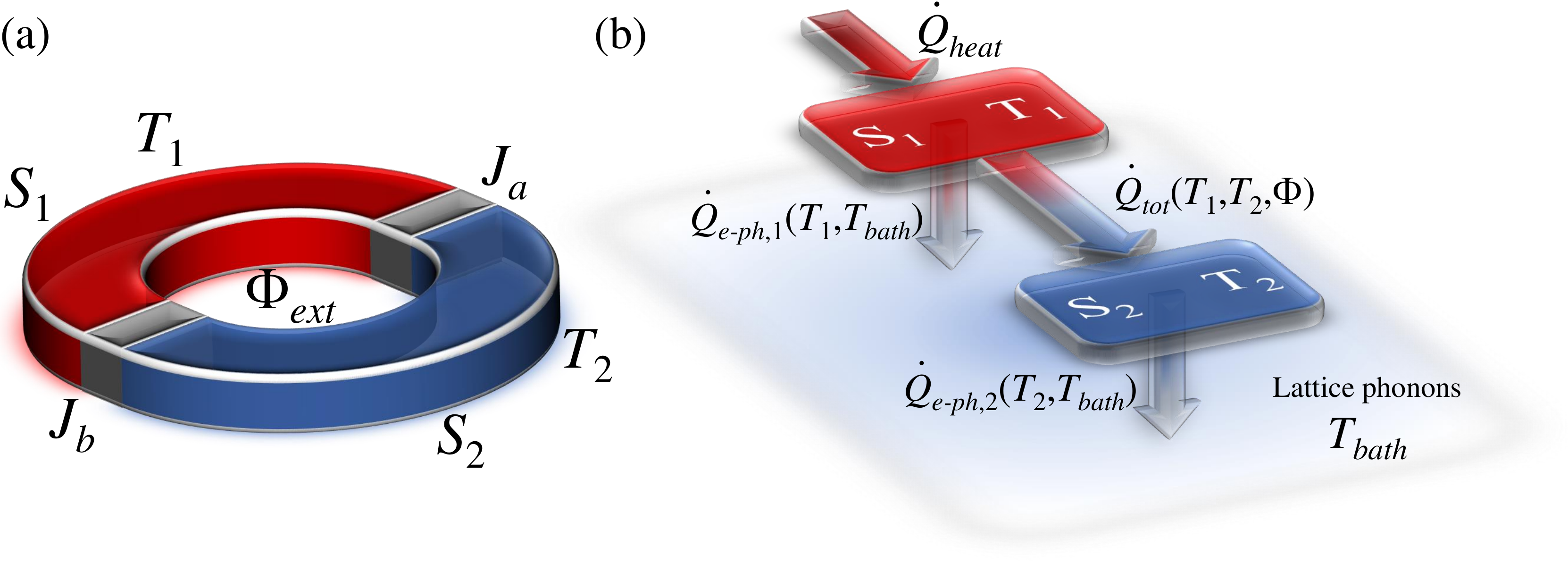}
\caption{(Color online) (a) Two superconductors $S_1$ and $S_2$ kept at temperature $T_1$ and $T_2$ (with $T_{1}\geq T_{2}$), respectively, are tunnel-coupled through the junctions $J_a$ and $J_b$ so to implement a DC SQUID. $\Phi_{ext}$ is the
applied magnetic flux threading the SQUID loop. (b)
Sketch of the thermal model accounting for heat transport in the system. The total heat current, $\dot{Q}_{tot}$, flowing in the system depends on the temperatures and the total flux $\Phi$ (see Eq.~\ref{TotalFlux}) through the SQUID ring. $\dot{Q}_{e-ph,i}(T_i,T_{bath})$ represents the coupling between quasiparticles in $S_i$ and the lattice phonons residing at $T_{bath}$, whereas $\dot{Q}_{heat}$ denotes the power injected into $S_1$ through heating probes in order to impose a quasiparticle temperature $T_1$. The arrows indicate the direction of heat currents for $T_{bath} < T_2 < T_1$.}
\label{Fig00}
\end{figure*}

The device we are discussing is a double-tunnel-junction superconducting quantum interference device, namely, a DC SQUID, formed by two superconductors $S_1$ and $S_2$ in a thermal steady state residing at different temperatures $T_1$ and $T_2$, respectively, with $T_1\geq T_2$ (see Fig~\ref{Fig00}a).
In the presence of a temperature gradient and with no voltage bias, a stationary finite heat current, $\dot{Q}_{tot}$, given by
\begin{equation}
 \dot{Q}_{tot} ( T_1,T_2)= \dot{Q}_{qp}( T_1,T_2)- \dot{Q}_{int}\left ( T_1,T_2,\varphi_a,\varphi_b \right ),
\label{TotalThermalCurrent}
\end{equation}
flows from $S_1$ to $S_2$~\cite{Mak65,Gut97,Gut98,Zha03,Zha04,GiaMar12,Gol13}, see Fig~\ref{Fig00}. Eq.~(\ref{TotalThermalCurrent}) contains the interplay between Cooper pairs and quasi-particles in tunneling through a JJ predicted by Maki and Griffin~\cite{Mak65}. In fact, the term
\begin{equation}
\dot{Q}_{qp}=\dot{Q}^a_{qp}( T_1,T_2)+\dot{Q}^b_{qp}( T_1,T_2)
\label{TotalQPCurrent}
\end{equation}
is the heat flux carried by quasiparticles and represents an incoherent flow of energy from the hot to the cold electrode~\cite{Mak65,Gia06,Fra97} through the junctions $J_{a}$ and $J_{b}$. Instead, the term
\begin{equation}
\dot{Q}_{int}=\dot{Q}^a_{int}( T_1,T_2)\cos\varphi_a+\dot{Q}^b_{int}( T_1,T_2)\cos\varphi_b
\label{TotalIntCurrent}
\end{equation}
is the phase-dependent part of the heat current~\cite{Mak65,Gut97,Zha03,Zha04,Gol13} ($\varphi_{a(b)}$ being the macroscopic quantum phase difference between the superconductors across the junction $J_{a(b)}$). It is peculiar to the tunnel JJs forming the SQUID and is the thermal counterpart of the ``quasiparticle-pair interference'' term contributing to the electrical current tunneling through a JJ~\cite{Bar82}. 
This term originates from the energy-carrying tunneling processes involving recombination and destruction of Cooper pairs on both sides of each junction, and is therefore governed by the phase difference $\varphi_{a(b)}$ between the two superconducting condensates. 
The oscillatory behavior of the thermal current was experimentally verified in Refs.~\cite{Gia12,Mar14,For16}.

The terms of Eq.~(\ref{TotalQPCurrent}) explicitly read~\cite{Mak65,Gut97,Gut98,Zha03,Zha04,Gol13}
\begin{small}
\begin{equation}\label{QqpSQUID}
\dot{Q}^{a(b)}_{qp}\text{=}\frac{1}{e^2R_{a(b)}}\int_{0}^{\infty}d\varepsilon \varepsilon \mathcal{N}_1 ( \varepsilon ,T_1 )\mathcal{N}_2 ( \varepsilon ,T_2 ) [ f ( \varepsilon ,T_2 ) -f ( \varepsilon ,T_1 ) ],
\end{equation}
\end{small}%
where
\begin{equation}
\mathcal{N}_i\left ( \varepsilon ,T_i \right )=\left |\text{Re}\left [\frac{ \varepsilon + i\Gamma_i}{\sqrt{\left (\varepsilon+ i\Gamma_i \right ) ^2-\Delta _i\left ( T_i \right )^2}} \right ] \right |
\end{equation}
is the smeared normalized BCS density of states in $S_i$ at temperature $T_i$ ($i=1,2$), $\Gamma_i$ being the Dynes parameter~\cite{Dyn78}. Hereafter, we set $\Gamma_i=10^{-4}\Delta_i(0)$, a value which describes realistic superconducting tunnel junctions~\cite{Mar15,ForTimBos16}. Here, $\varepsilon$ is the energy measured from the condensate chemical potential, $\Delta _i\left ( T_i \right )$ is the temperature-dependent superconducting energy gap, $f\left ( \varepsilon ,T_i \right )=\tanh\left ( \varepsilon/2 k_B T_i \right )$, $R_{a(b)}$ is the junction normal-state resistance, $k_B$ is the Boltzmann constant, and $e$ is the electron charge. The terms of Eq.~(\ref{TotalIntCurrent}) read~\cite{Mak65,Gut97,Gut98,Zha03,Zha04,Gol13}
\begin{small}
\begin{equation}\label{QintSQUID}
\dot{Q}^{a(b)}_{int}\text{=}\frac{1}{e^2R_{a(b)}}\int_{0}^{\infty}d\varepsilon \varepsilon \mathcal{M}_1 ( \varepsilon ,T_1 )\mathcal{M}_2 ( \varepsilon ,T_2) [ f ( \varepsilon ,T_2 ) - f ( \varepsilon ,T_1 ) ],
\end{equation}
\end{small}%
where
\begin{equation}
\mathcal{M}_i\left ( \varepsilon ,T_i \right )=\left |\text{Im}\left [\frac{ - i\Delta _i\left ( T_i \right )}{\sqrt{\left (\varepsilon+ i\Gamma_i \right ) ^2-\Delta _i\left ( T_i \right )^2}} \right ] \right |
\end{equation}
is the Cooper pair BCS density of states in $S_i$ at temperature $T_i$~\cite{Bar82}.
We note that both $\dot{Q}^{a(b)}_{qp}$ and $\dot{Q}^{a(b)}_{int}$ vanish for $T_1=T_2$, while $\dot{Q}^{a(b)}_{int}$ also vanishes when at least one of the superconductors is in the normal state, i.e., $\Delta _i\left ( T_i \right )=0$.

According to the conservation of the supercurrent circulating in the loop, the phases $\varphi_{a}$ and $\varphi_{b}$ satisfy the equation
\begin{equation}
I=I_J^a\sin\varphi_a=I_J^b\sin\varphi_b.
\label{supercurrents}
\end{equation}
Here, $I_J^{a(b)}( T_1,T_2)$ is the critical current for the temperature-biased junction $J_{a(b)}$, given by~\cite{Gia05,Tir08,Bos16}
\begin{eqnarray}\label{IcT1T2}\nonumber
I_J^{a(b)} ( T_1,T_2 )\textup{=}&&\frac{1}{2eR_{a(b)}}\Bigg|\int_{-\infty}^{\infty} \Big\{ f( \varepsilon ,T_1 )\textup{Re}\left [\mathfrak{F}_1(\varepsilon ) \right ]\textup{Im}\left [\mathfrak{F}_2(\varepsilon ) \right ]\\
&&+ f( \varepsilon ,T_2 )\textup{Re}\left [\mathfrak{F}_2(\varepsilon ) \right ]\textup{Im}\left [\mathfrak{F}_1(\varepsilon ) \right ] \Big\} d\varepsilon \Bigg|,
\end{eqnarray}
where $\mathfrak{F}_j(\varepsilon ) =\Delta_j \left ( T_j \right )\Big/\sqrt{\left ( \varepsilon +i\Gamma_j \right )^2-\Delta_j^2 \left ( T_j\right )}$.
In the following we assume $\Delta_1(0)=\Delta_2(0)=\Delta=1.746k_BT_c$, $T_c$ being the common critical temperature of the superconductors. An in-plane external magnetic field also causes the JJs critical current to modulate~\cite{Bar82}. However, since the area of the SQUID ring is usually greater than the area of the junctions, this modulation occurs on a field scale much larger than the modulation period of thermal currents in the SQUID. Moreover, JJs in the so-called overlap geometry are usually preferred in SQUID-based applications~\cite{Mar14}, so that the magnetic field threading the loop is out-of-plane to the junctions area and no modulation of the critical currents occurs. Therefore, hereafter the junction critical currents $I_J$ are assumed independent of the external flux variations. 

The degree of asymmetry of the SQUID, $\alpha$, is defined as the critical currents ratio, so that
\begin{equation}
\alpha =\frac{I_J^a}{I_J^b}=\frac{R_b}{R_a}=\frac{\dot{Q}^{a}_{qp}}{\dot{Q}^{b}_{qp}}=\frac{\dot{Q}^{a}_{int}}{\dot{Q}^{b}_{int}},
\label{AsymmetryParameter}
\end{equation}
according to Eqs.~(\ref{QqpSQUID}),~(\ref{QintSQUID}), and~(\ref{IcT1T2}).
The flux quantization imposes the constraint
\begin{figure}[t!!]
\centering
\includegraphics[height=6.cm]{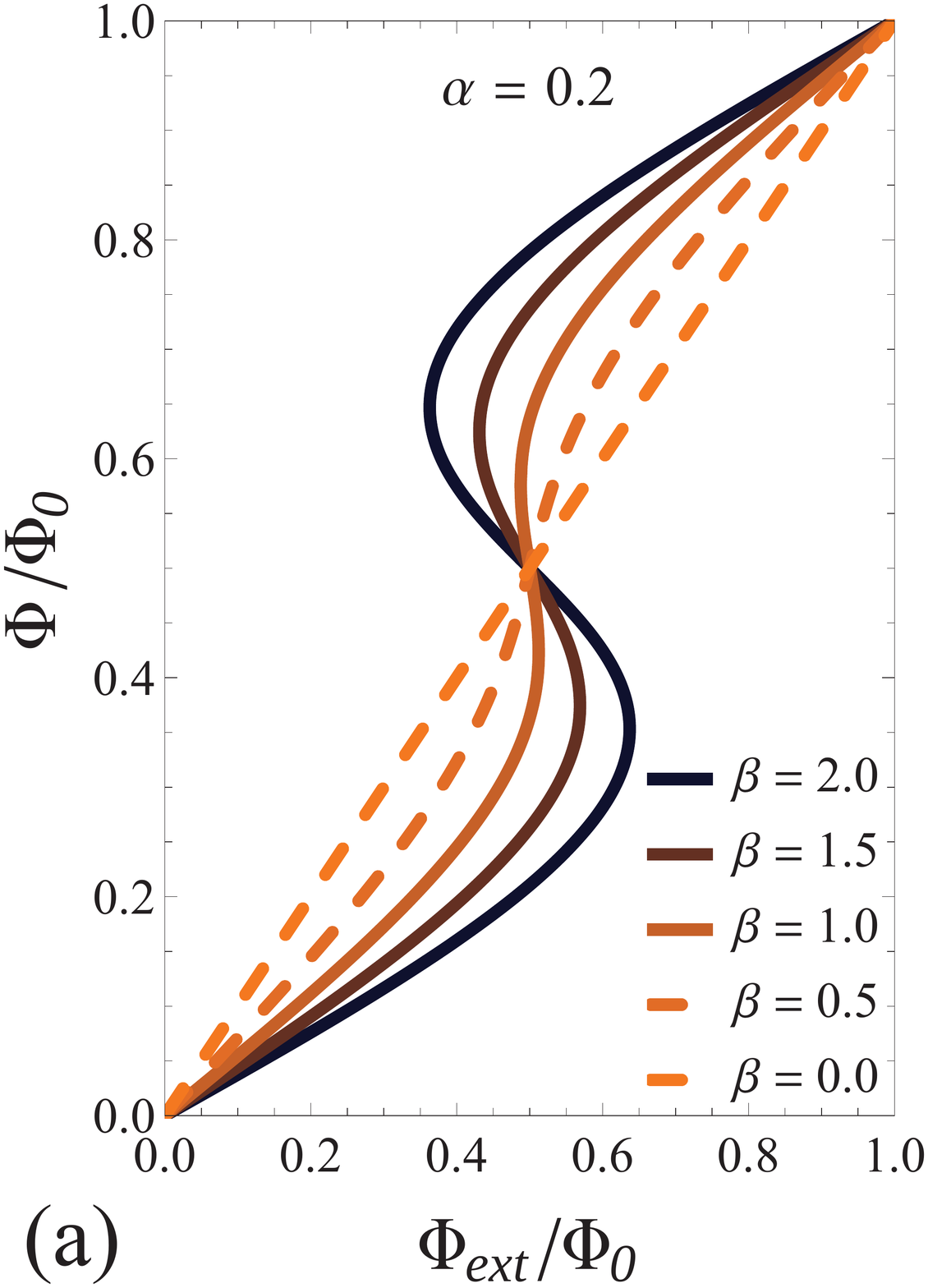}
\includegraphics[height=6.cm]{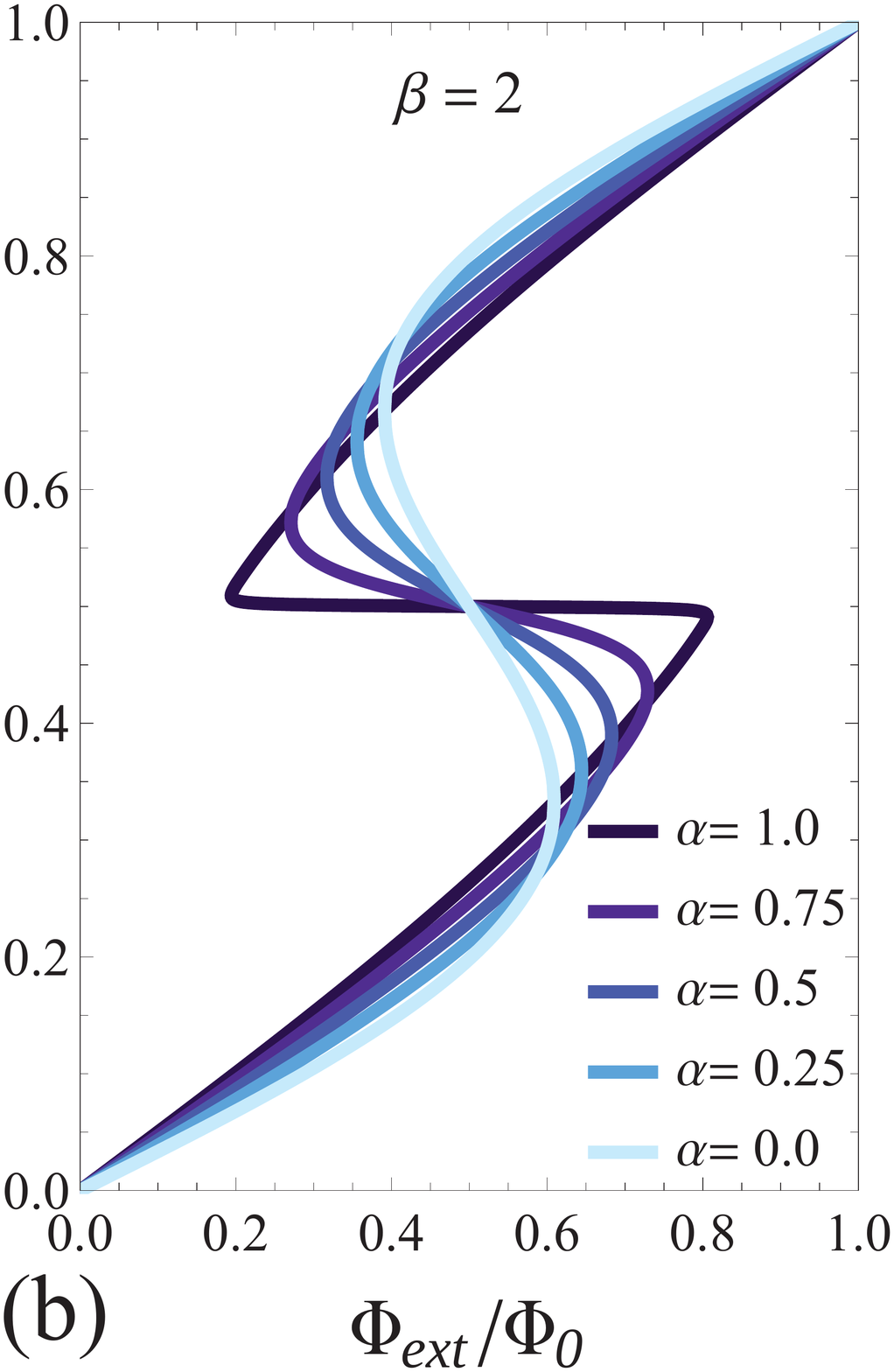}
\caption{(Color online) Normalized total magnetic flux $\Phi/\Phi_0$ as a function of the normalized external magnetic flux $\Phi_{ext}/\Phi_0$ for $\alpha=0.2$ and several values of $\beta$, see panel (a), and for $\beta=2$ and several values of $\alpha$, see panel (b). Dashed curves represent non-hysteretic conditions.}
\label{Fig01}
\end{figure}
\begin{equation}
\varphi_a+\varphi_b+2\pi \frac{\Phi}{\Phi_0}=2\pi k
\label{Fluxquantization}
\end{equation}
where $\Phi _0\simeq 2.067\times 10^{-15}\;\textup{Wb}$ is the flux quantum and $k$ is an integer representing the amount of enclosed flux quanta, so that the transition $k\to k\pm 1$ indicates a variation of one flux quantum through the SQUID ring. 
In Eq.~(\ref{Fluxquantization}), $\Phi$ is the total magnetic flux given by
\begin{equation}
\Phi =\Phi_{ext}+LI,
\label{TotalFlux}
\end{equation}
where $\Phi_{ext}$ is the externally applied magnetic flux through the ring, see Fig~\ref{Fig00}a, and the SQUID inductance $L$ has a geometric contribution as well as a kinetic contribution~\cite{Maj02,Cla04}. 
From Eqs.~(\ref{supercurrents}),~(\ref{AsymmetryParameter}), and~(\ref{Fluxquantization}), one obtains the circulating current, in units of $I_J^a$, as a function of the total flux~\cite{Bo04}
\begin{equation}
\frac{I}{I_J^a}=\sin\varphi_a=\frac{-\sin\left ( 2\pi \frac{\Phi }{\Phi_0} \right )}{\sqrt{1+\alpha^2+2\alpha \cos\left ( 2\pi \frac{\Phi }{\Phi_0} \right )}},
\label{NormSupercurrent}
\end{equation}
so that, from Eq.~(\ref{TotalFlux}) we obtain
\begin{equation}
\frac{\Phi}{\Phi_0} =\frac{\Phi_{ext}}{\Phi_0}-\frac{\beta\sin\left ( 2\pi \frac{\Phi }{\Phi_0} \right )}{2\pi\sqrt{1+\alpha^2+2\alpha \cos\left ( 2\pi \frac{\Phi }{\Phi_0} \right )}}, 
\label{NormTotalFlux}
\end{equation}
$\beta=2\pi L I_J^a/\Phi_0$ being the hysteresis parameter. 
\begin{figure}[t!!]
\centering
\includegraphics[width=\columnwidth]{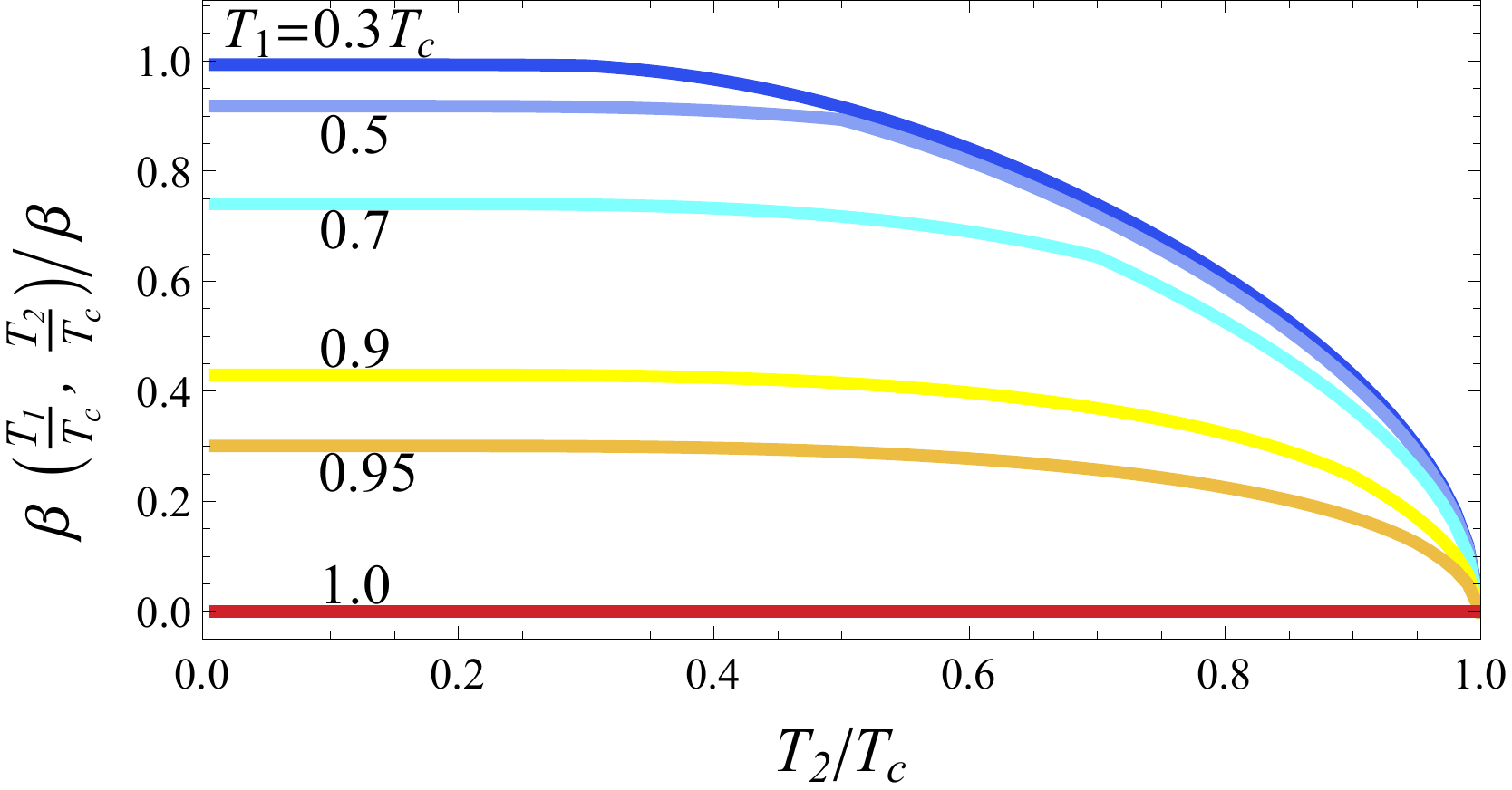}
\caption{(Color online) Hysteresis parameter $\beta(T_1/T_c,T_2/T_c)$, in units of $\beta\equiv\beta(0,0)$, as a function of the normalized temperature $T_2/T_c$, for a few values of $T_1/T_c$.}
\label{Fig04}
\end{figure}

Before exploring the behavior of the total heat current by changing the temperatures $T_1$ and $T_2$ of the electrodes, we observe that the hysteretic parameter $\beta$ depends on them through the critical current $I_J^a$ according to Eq.~(\ref{IcT1T2}). As is clearly shown in Fig.~\ref{Fig04}, $\beta(T_1,T_2)\to0$ as the temperatures approach $T_c$, and so the hysteresis reduces by increasing the temperature. In the following, a temperature-dependent hysteretic parameter is taken into account and the notation $\beta\equiv\beta(0,0)$ is used.

When $\alpha\to0$ (i.e., single junction SQUID) and $\alpha\to1$ (i.e., symmetric SQUID), Eq.~(\ref{NormTotalFlux}) turns into $\Phi=\Phi_{ext} -LI_J^a\sin(2\pi\Phi/\Phi_0)$ and $\Phi=\Phi_{ext} -LI_J^a\sin(\pi\Phi/\Phi_0)$, respectively.

For proper values of $\alpha$ and $\beta$ the total flux $\Phi$ is a multi-valued function of $\Phi_{ext}$ and the SQUID behaves hysteretically~\cite{Bo04}. Specifically, for $\beta<1-\alpha$ the slope of $\Phi$ is always positive and the $\Phi\;vs\;\Phi_{ext}$ plot is non-hysteretic (see dashed curves in Fig.~\ref{Fig01}a). Conversely, for $\beta>1-\alpha$ the slope of $\Phi$ switches from positive to negative, so that $\Phi\;vs\;\Phi_{ext}$ is multi-valued and a hysteretic curve results (see Fig.~\ref{Fig01}a). Moreover, by increasing the values of $\beta$ and $\alpha$ the range of $\Phi_{ext}$ values in which $\Phi$ has negative slopes enlarges, and, accordingly, the hysteresis of the $\Phi\;vs\;\Phi_{ext}$ curves is more pronounced, see Fig.~\ref{Fig01}a and Fig.~\ref{Fig01}b, respectively. 

\begin{figure}[t!!]
\centering
\includegraphics[width=\columnwidth]{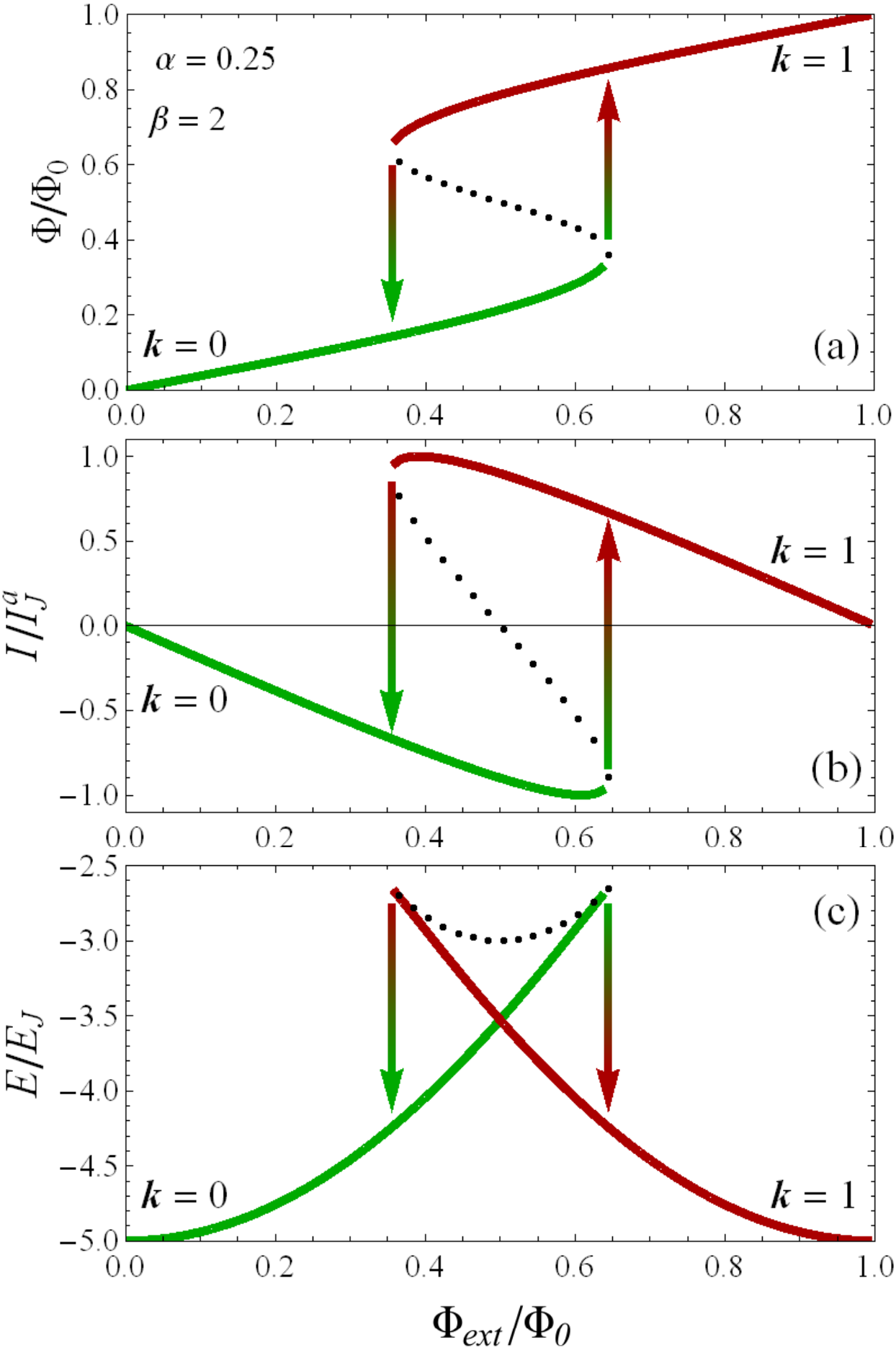}
\caption{(Color online) Normalized total flux, normalized supercurrent, and free energy of the SQUID as a function of the external flux $\Phi_{ext}/\Phi_0$, see panels (a), (b), and (c), respectively, for $\alpha=0.25$ and $\beta=2$. Arrows indicate the transition between quantum states with different number $k$ of flux quanta penetrating the loop. Dot curves represent the unstable states in the hysteretic mode.}
\label{Fig02}
\end{figure}
The adiabatic evolution of the system is obtained from the minimization of the free energy of the SQUID, that is composed of the Josephson energies and the inductive energy, due to the screening current flowing into the SQUID ring,
\begin{equation}
E=-\frac{\Phi_0}{2\pi}\left ( I_J^a\cos\varphi_a + I_J^b\cos\varphi_b\right )+\frac{1}{2}LI^2.
\label{FreeEnergy}
\end{equation}
\begin{figure}[t!!]
\centering
\includegraphics[height=6.cm]{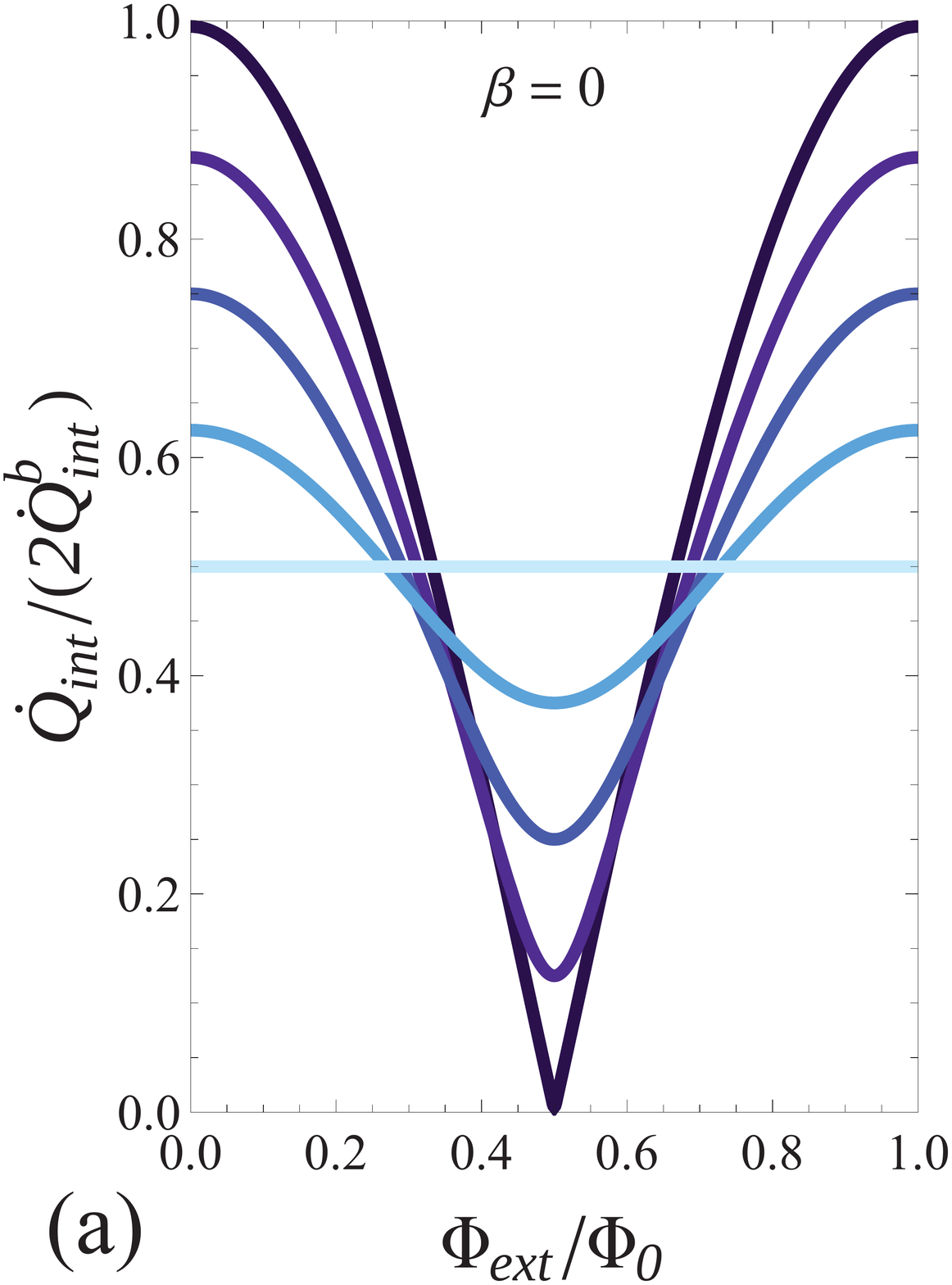}
\includegraphics[height=6.cm]{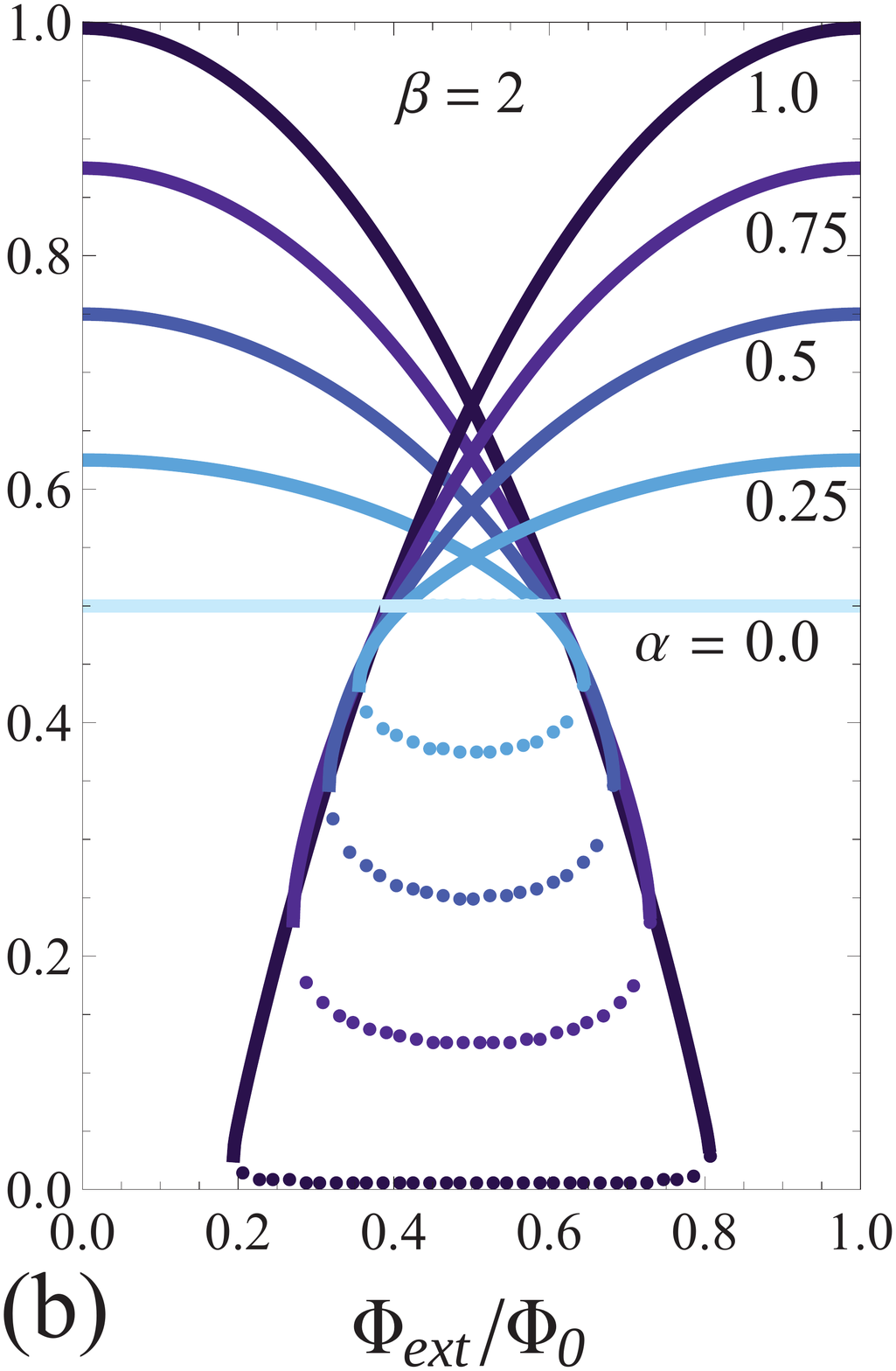}
\caption{(Color online) Interference heat current $\dot{Q}_{int}$ as a function of the normalized external magnetic flux $\Phi_{ext}/\Phi_0$, for several values of $\alpha$, for $\beta=0.0$ and $\beta=2$, see panels (a) and (b), respectively. Here, we set generic temperatures $T_1$ and $T_2$, such that $T_2<T_1<T_c$. Dot curves represent the unstable states in the hysteretic mode.}
\label{Fig03}
\end{figure}
From Eqs.~(\ref{supercurrents}) and~(\ref{NormSupercurrent}), the free energy (in units of $E_J=\frac{\Phi_0}{2\pi}I_J^a$) becomes
\begin{eqnarray}\nonumber
\frac{E}{E_J}=&-&\frac{1}{\alpha}\sqrt{1+\alpha^2+2\alpha\cos\left (2\pi \frac{\Phi}{\Phi_0} \right )}+	\\	
&+&\frac{\beta}{2}\frac{\sin^2\left (2\pi \frac{\Phi}{\Phi_0} \right )}{1+\alpha^2+2\alpha\cos\left (2\pi \frac{\Phi}{\Phi_0} \right )}.
\label{SQUIDFreeEnergy}
\end{eqnarray}
%
%
%
The hysteretic behaviors of the total flux, the normalized Josephson current, and the free energy of a SQUID with $\alpha=0.25$ and $\beta=2$, are shown in panels (a), (b), and (c) of Fig.~\ref{Fig02}, respectively. 
The effect of the inductance is evident in Fig.~\ref{Fig02}a where the total flux $\Phi$ grows less rapidly than $\Phi_{ext}$, for the flux generated by the screening current opposes $\Phi_{ext}$. 
As $\left | I \right |$ exceeds the critical value $I_J^a$ (see Fig.~\ref{Fig02}b), the junction temporarily switches into the voltage state~\cite{Bar82}. Correspondingly, a jump to a lower free energy occurs (see Fig.~\ref{Fig02}c) and the SQUID undergoes a quantum transition $k\to k+1$, so that the flux through the SQUID changes by one flux quantum (see Fig.~\ref{Fig02}a). Further reducing the external flux, the system remains in the $k=1$ state until the circulating current $\left | I \right |$ reaches the critical value $I_J^a$ and the free energy jumps again to a lower value, when the SQUID switches to the $k=0$ state. We note that for negative slopes of $\Phi$ (dotted curves in Fig.~\ref{Fig02}) the SQUID free energy is definitively higher with respect to the energies of the states for positive $\Phi$ slopes. Therefore, the states corresponding to negative $\Phi$ slopes (dot curves in Fig.~\ref{Fig02}) are definitively unstable and are not observed during an adiabatic evolution, so that in sweeping back and forth $\Phi_{ext}$ a hysteretic path is traced out. Hereafter, dot curves in the figures represent unstable states of the SQUID.

According to Eqs.~(\ref{supercurrents}) and~(\ref{NormSupercurrent}), one gets~\cite{Bo04}
\begin{eqnarray}\label{CosPhia}
\cos\varphi_a&=&\frac{\alpha+\cos\left ( 2\pi\frac{\Phi}{\Phi_0} \right )}{\sqrt{1+\alpha^2+2\alpha\cos\left ( 2\pi\frac{\Phi}{\Phi_0} \right )}}\\
\cos\varphi_b&=&\frac{1+\alpha\cos\left ( 2\pi\frac{\Phi}{\Phi_0} \right )}{\sqrt{1+\alpha^2+2\alpha\cos\left ( 2\pi\frac{\Phi}{\Phi_0} \right )}},
\label{CosPhib}
\end{eqnarray}
so that $\dot{Q}_{int}$, see Eq.~(\ref{TotalIntCurrent}), becomes~\cite{GiaMar12}
\begin{equation}
\dot{Q}_{int}=\dot{Q}^b_{int}\left ( T_1,T_2 \right )\sqrt{1+\alpha^2+2\alpha\cos\left ( 2\pi\frac{\Phi}{\Phi_0} \right )}.
\label{TotalIntCurrentSQUID}
\end{equation}
The behavior of the interference heat current $\dot{Q}_{int}$ as a function of the external flux, for a few values of $\alpha$ and for $\beta=0.0$ and $\beta=2$ is shown in panels (a) and (b) of Fig.~\ref{Fig03}, respectively. Generic temperatures $T_1$ and $T_2$, such that $T_2<T_1<T_c$, are set. $\dot{Q}_{int}$ is a periodic function of the external flux, and it is modulated between the maximum, given by $\dot{Q}^{^M}_{int}=\dot{Q}^b_{int}(1+\alpha)$ for $\Phi_{ext}=n\Phi_0$, and the minimum value, given by $\dot{Q}^{^m}_{int}=\dot{Q}^b_{int}(1-\alpha)$ for $\Phi_{ext}=(n+1/2)\Phi_0$ ($n$ is an integer). Therefore, the modulation amplitude of $\dot{Q}_{int}$ is totally suppressed for $\alpha=0$, so that high junction symmetry is required to maximize the heat current modulation in the device. 
By increasing the hysteresis parameter $\beta$, $\dot{Q}_{int}$ is multivalued in a neighbourhood of $\Phi_0/2$, so that the dip in $\Phi_{ext}=\Phi_0/2$ for $\beta=0.0$ is replaced by a loop, whose width increases with increasing both $\beta$ and $\alpha$. We observe that the bottom of these loops (see dot curves in Fig.~\ref{Fig03}b) corresponds to the unstable states of the SQUID. We also note that the height of the hysteretic jumps increases for $\alpha\to1$.

\begin{figure}[t!!]
\centering
\includegraphics[width=\columnwidth]{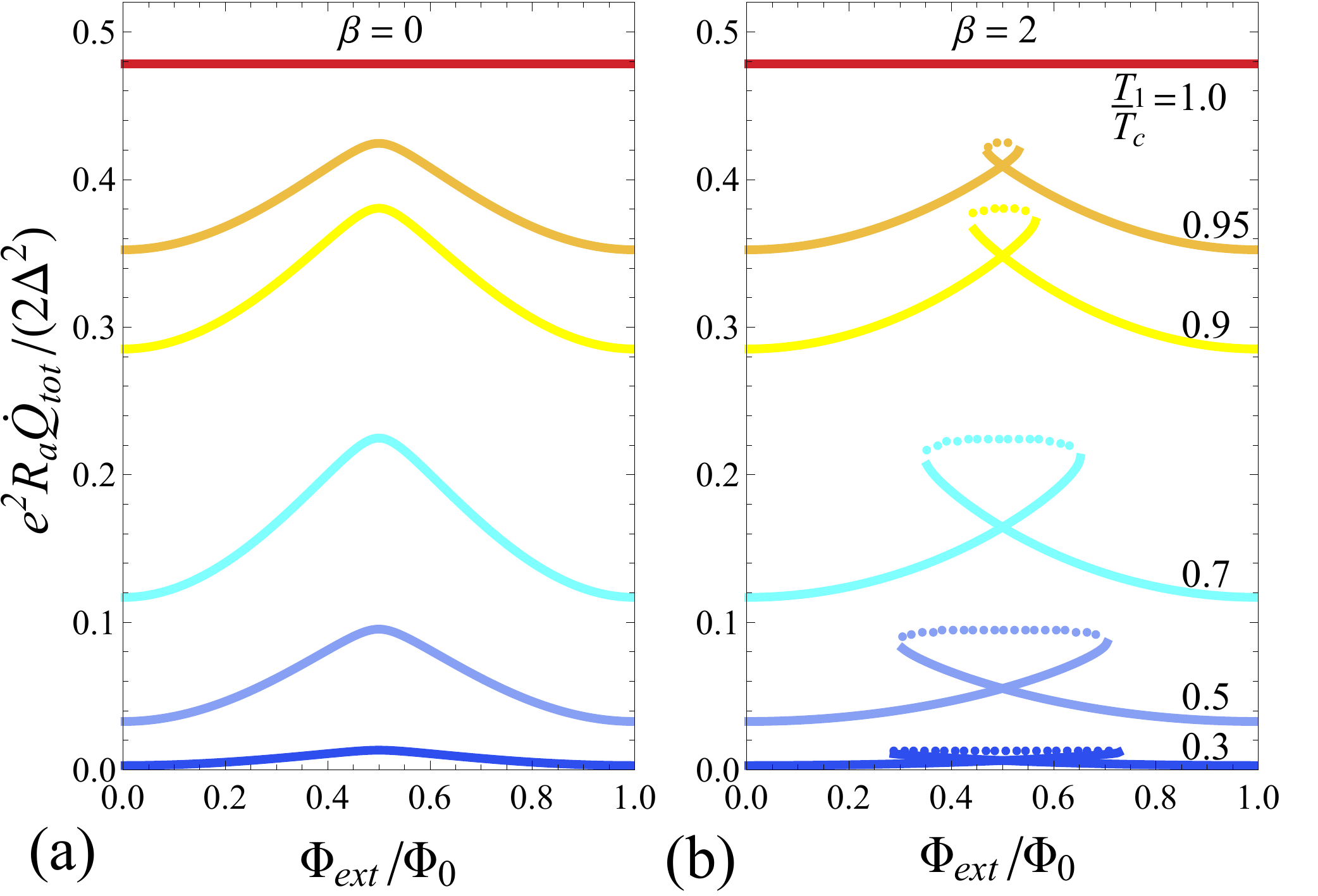}
\caption{(Color online) Total heat current $\dot{Q}_{tot}$ as a function of the normalized external magnetic flux $\Phi_{ext}/\Phi_0$ for a few values of $T_1$ and $T_2=0.1T_c$, assuming $\alpha=0.75$, for $\beta=0.0$ and $\beta=2$, see panels (a) and (b), respectively. Dotted curves represent the unstable states in the hysteretic mode.
}
\label{Fig05}
\end{figure}

\begin{figure}[t!!]
\centering
\includegraphics[width=\columnwidth]{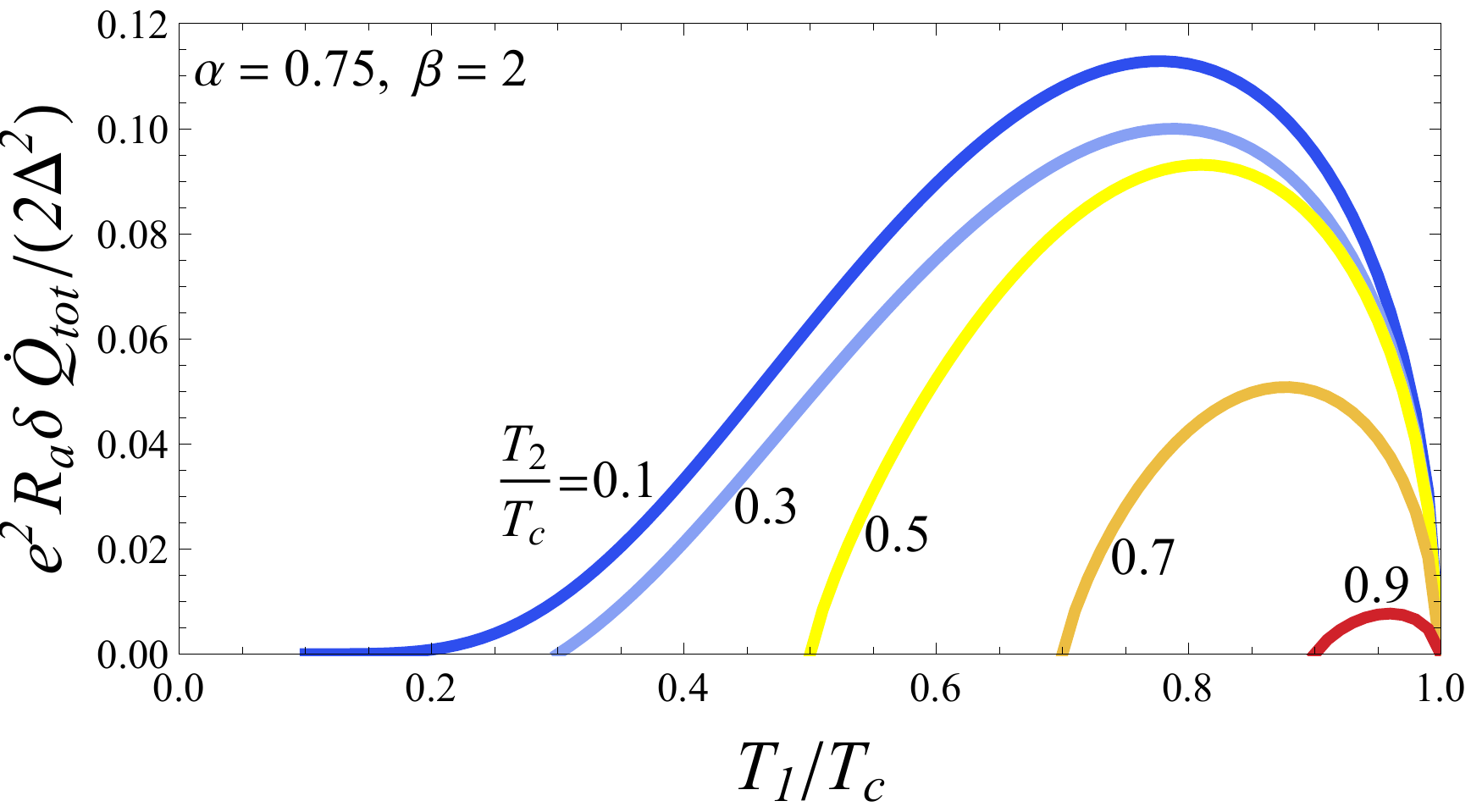}\\
\includegraphics[width=\columnwidth]{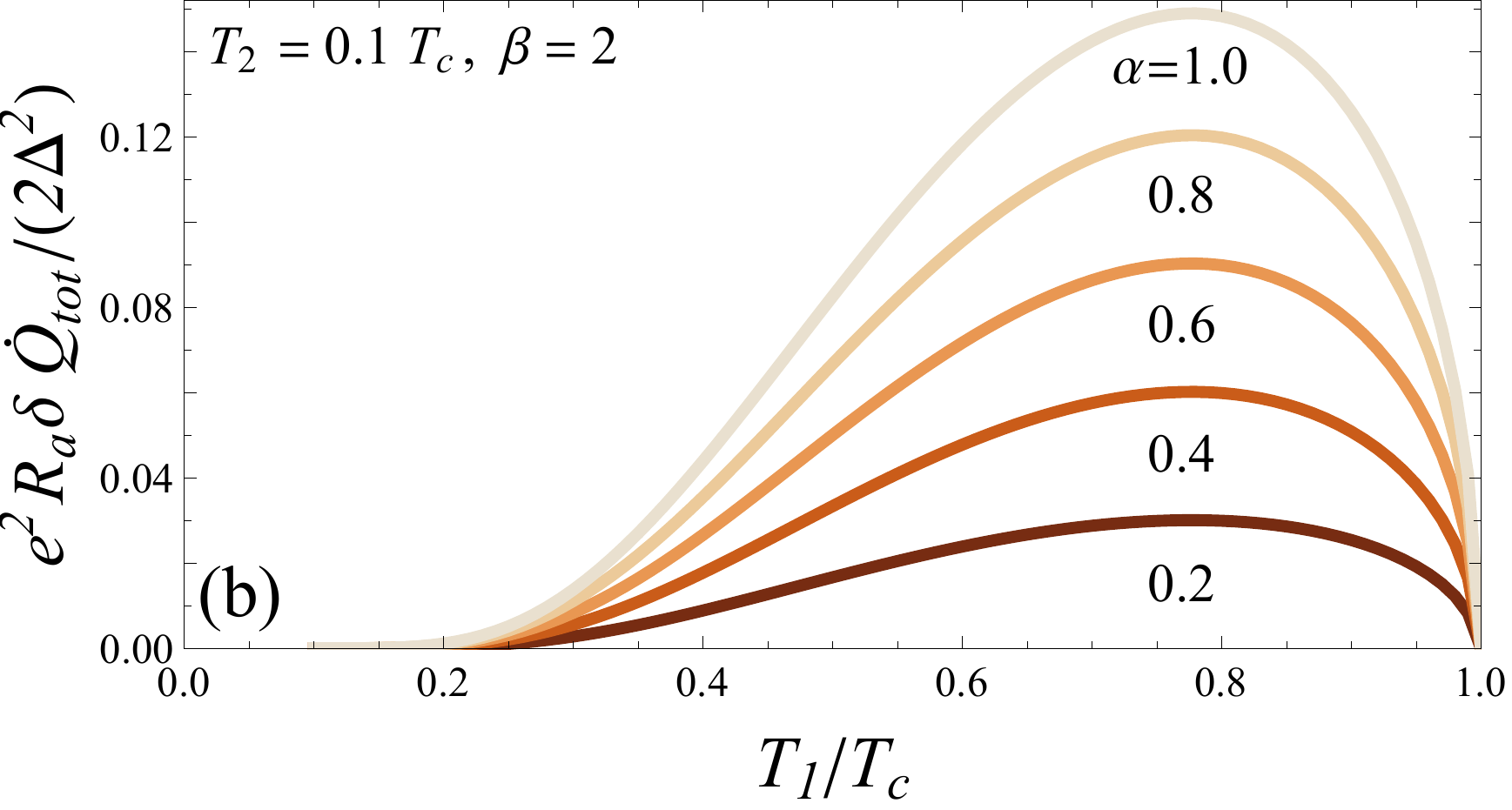}
\caption{(Color online) Total heat current modulation amplitude $\delta\dot{Q}_{tot}$ as a function of the normalized temperature $T_1/T_c$, assuming $\beta=2$, for a few values of $T_2$ and $\alpha=0.75$, panel (a), and for a few values of $\alpha$ and $T_2=0.1T_c$, panel (b). }
\label{Fig06}
\end{figure}

Fig.~\ref{Fig05} shows the total heat current $\dot{Q}_{tot}$ as a function of $\Phi_{ext}$ at $T_2=0.1T_c$ for several values of $T_1$, for $\beta=0.0$ and $\beta=2$, see panel (a) and (b), respectively. As expected, $\dot{Q}_{tot}(\Phi_{ext})$ is modulated with the same $\Phi_0$-periodicity of $\dot{Q}_{int}$, but, unlike the latter, it is minimized (maximized) for integer (half-integer) values of $\Phi_0$, according to a minus sign in front of the $\varphi$-dependent term in Eq.~(\ref{TotalThermalCurrent}). Similar to $\dot{Q}_{int}$, the hysteretic loop for $\beta>0$ appears also in $\dot{Q}_{tot}$, although the temperature-dependence of the hysteresis parameter $\beta$ makes the curves less and less hysteretic as $T_1\to T_c$ (see Fig.~\ref{Fig05}b). 
We observe that at the lowest $T_1$, the total heat current is small, for $\dot{Q}_{int}$ and $\dot{Q}_{qp}$ are almost comparable. The total heat current modulation amplitude, $\delta\dot{Q}_{tot}$, defined as the difference between the maximum and the minimum values of $\dot{Q}_{tot}$, reduces further by increasing $T_1$, and it vanishes for $T_1=T_c$ when $S_1$ is driven into the normal state. In Fig.~\ref{Fig06}a the behavior of $\delta\dot{Q}_{tot}$ is shown, for the values of $\alpha$, $\beta$, and $T_2$ used to obtain data in Fig.~\ref{Fig05}b. Specifically, $\delta\dot{Q}_{tot}$ is a non-monotonic function vanishing for $T_1=T_c$, with a maximum in correspondence to an intermediate temperature depending on $T_2$. In particular, by increasing $T_2$ the maximum value of $\delta\dot{Q}_{tot}$ reduces and shifts towards higher $T_1$. Panel b of Fig.~\ref{Fig06} shows the modulation amplitude $\delta\dot{Q}_{tot}$ at $T_2=0.1T_c$ for a few values of the asymmetry parameter $\alpha$. Here, it is clearly shown that the modulation reduces with $\alpha$, whereas the position of the maximum is not affected by $\alpha$ variations.

\section{Thermal Model}
\label{ThermalModel}\vskip-0.2cm 

\begin{figure}[t!!]
\centering
\includegraphics[width=\columnwidth]{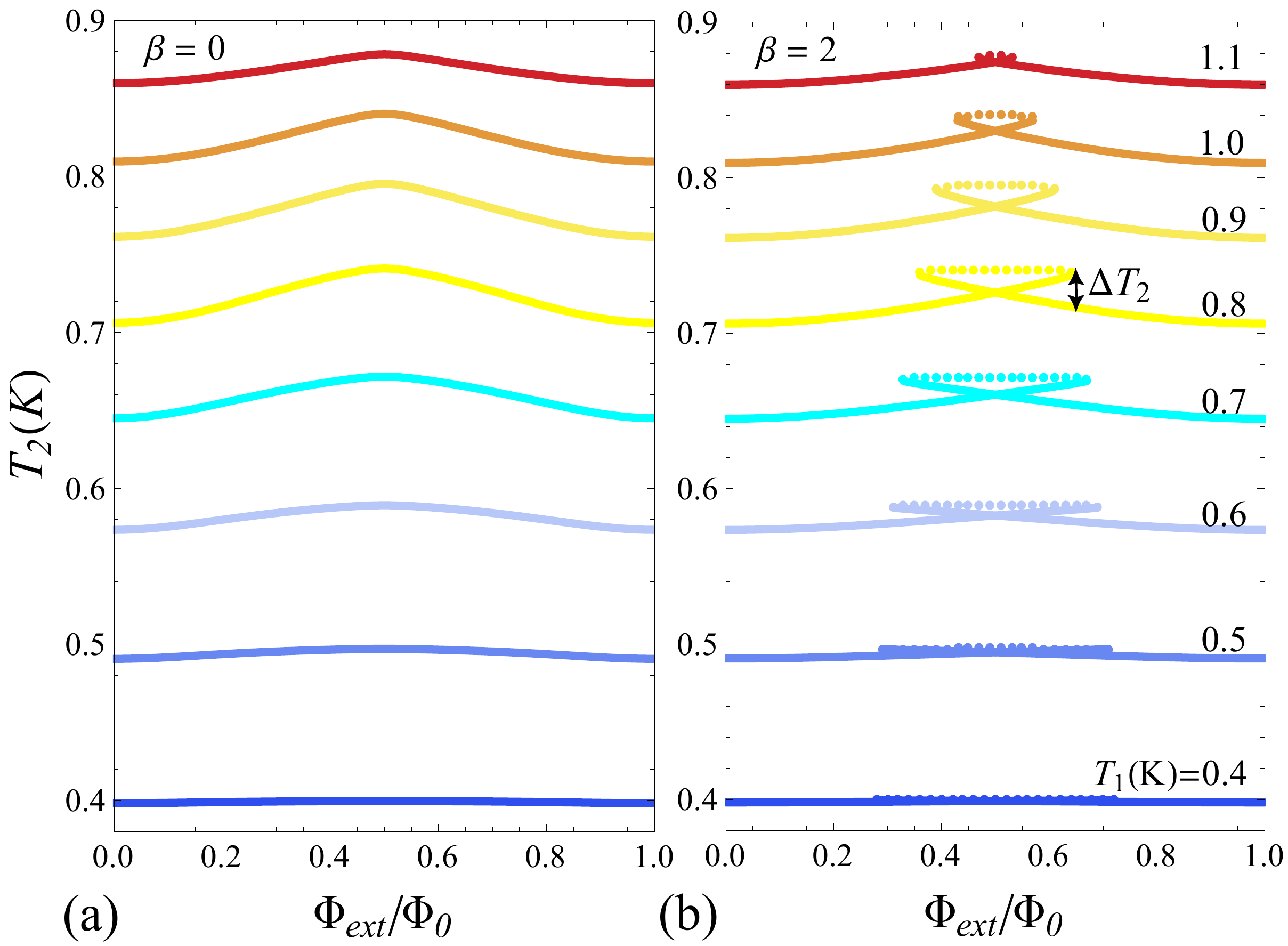}
\caption{(Color online) Quasiparticle temperature $T_2\;vs\;\Phi_{ext}$ calculated for a few values of $T_1$ at $T_{bath} = 0.1 \text{K}$ for $\alpha=0.75$, $\beta=0.0$ and $\beta=2$, panel (a) and (b), respectively. Dotted curves represent the unstable states in the hysteretic mode. In panel b the height of a $T_2$ jump, $\Delta T_2$, as $\Phi_{ext}$ induces a transition in the number of flux quanta through the SQUID ring, $k\to k\pm1$, is indicated.}
\label{Fig07}
\end{figure}

A practical experimental setup to observe thermal modulation was proposed in Ref.~\cite{GiaMar12} and successfully implemented in Ref.~\cite{Gia12}. The device consists of a tunnel-junction DC SQUID formed by identical superconductors, driven by a magnetic flux. Superconducting leads tunnel-coupled to both SQUID electrodes, and serving either as heaters or thermometers (not shown in Fig.~\ref{Fig00}), allow to perturb and to accurately probe the quasiparticle temperature in the structure~\cite{Gia06}. The superconducting JJs provide nearly ideal thermal isolation of the SQUID electrodes~\cite{Gia06} and, therefore, the thermal conductance through these probes can be neglected. 
The thermal model that we are going to discuss is sketched in Fig.~\ref{Fig00}b. The steady temperature $T_2(\Phi)$ depends on the energy relaxation mechanisms occurring in the electrode $S_2$. For any $T_1$, the thermal balance equation for the incoming and outgoing thermal currents in $S_2$ can be written as~\cite{GiaMar12}
\begin{equation}
\dot{Q}_{tot}\left ( T_1,T_2,\Phi \right )-\dot{Q}_{e-ph,2}\left ( T_2,T_{bath}\right )=0.
\label{ThermalBalanceEq}
\end{equation}
Here, the heat current flowing from $S_1$ is balanced by the electron-phonon interaction~\cite{Gia06}, $\dot{Q}_{e-ph,2}$, namely the predominant energy relaxation mechanism in metals which allows energy exchange between quasiparticles and phonon bath. Specifically, $\dot{Q}_{e-ph,i}$ in $S_i$ reads~\cite{Pek09}
\begin{eqnarray}\label{Qe-ph}\nonumber
\dot{Q}_{e-ph,i}&\textup{=}&\frac{-\Sigma \mathcal{V}_i }{96\zeta(5)k_b^5}\int_{-\infty }^{\infty}dEE\int_{-\infty }^{\infty}d\varepsilon \varepsilon^2\textup{sign}(\varepsilon)M_{_{E,E+\varepsilon}}\times\\\nonumber
&\times& \Bigg\{ \coth\left ( \frac{\varepsilon }{2k_bT_{bath}}\right ) \left [ f(E,T_i)-f(E+\varepsilon,T_i) \right ]+\\
&-&f(E,T_i)f(E+\varepsilon,T_i)+1 \Bigg\} 
\end{eqnarray}
where $M_{E,{E}'}=\mathcal{N}_i(E,T_i)\mathcal{N}_i({E}',T_i)\left [ 1-\Delta ^2(T_i)/(E{E}') \right ]$, $\Sigma$ is the electron-phonon coupling constant, $\mathcal{V}_i$ is the volume of $S_i$, and $\zeta$ is the Riemann zeta function. In the following calculations, an aluminium DC SQUID with bulk critical temperature $T_c=1.19 \textup{K}$, $R_a=1\textup{k}\Omega$, $\mathcal{V}_2=10^{-19}\textup{m}^3$, and $\Sigma=3\times10^8\textup{Wm}^{-3}\textup{K}^{-5}$ is taken into account.

The temperature $T_2$ of the electrode $S_2$ is obtained by solving Eq.~(\ref{ThermalBalanceEq}). The behaviour of $T_2$ by varying the flux $\Phi_{ext}$ for several temperatures $T_1$ at a fixed bath temperature $T_{bath}=0.1\textup{K}$ is shown in Fig.~\ref{Fig07}, for $\beta=0.0$ and for $\beta=2$, see panels (a) and (b), respectively. The magnetic flux $\Phi_{ext}$ enclosed inside the SQUID ring modulates $T_2$ periodically, with a period of one flux quantum. By increasing $T_1$ the mean value of $T_2$ over a period increases. Markedly, the $T_2$ modulation amplitude, $\delta T_2$, defined as the difference between the maximum and the minimum value of $T_2$, behaves non-monotonically by varying $T_1$. In fact, $\delta T_2$ is vanishing for low $T_1$ (specifically, for $T_1=T_{bath}$ there is no thermal gradient along the system), then it increases up to $\delta T_2\sim35\textup{mK}$ for $T_1=0.84\textup{K}$ (see Fig.~\ref{Fig07}a), and finally it reduces again for $T_1\to T_c$, due to the temperature-induced suppression of the energy gaps in the superconductors. The flux modulation of $T_2$ in the hysteretic mode for $\beta=2$ is shown in Fig.~\ref{Fig07}b. The hysteretic behavior of the temperature $T_2$ as the flux $\Phi_{ext}$ is changed reflects the behavior of the total thermal current $\dot Q_{tot}$, although the temperature-dependence of $\beta$ makes the curves clearly less hysteretic as $T_1$ approaches $T_c$. The height of the $T_2$ jumps, $\Delta T_2$, as $\Phi_{ext}$ induces a transition in the number of flux quanta through the SQUID ring, $k\to k\pm1$, enhances by increasing $\alpha$ and $\beta$. Instead, for fixed $\alpha$ and $\beta$, it behaves non-monotonically by varying $T_1$, just like $\delta T_2$ in the non-hysteretic mode. In fact, $\Delta T_2$ vanishes for low $T_1$ and for $T_1\to T_c$, whereas it has a maximum $\Delta T_2\sim20\textup{mK}$ for $T_1=0.84\textup{K}$ (see Fig.~\ref{Fig07}b).

\begin{figure}[t!!]
\centering
\includegraphics[width=\columnwidth]{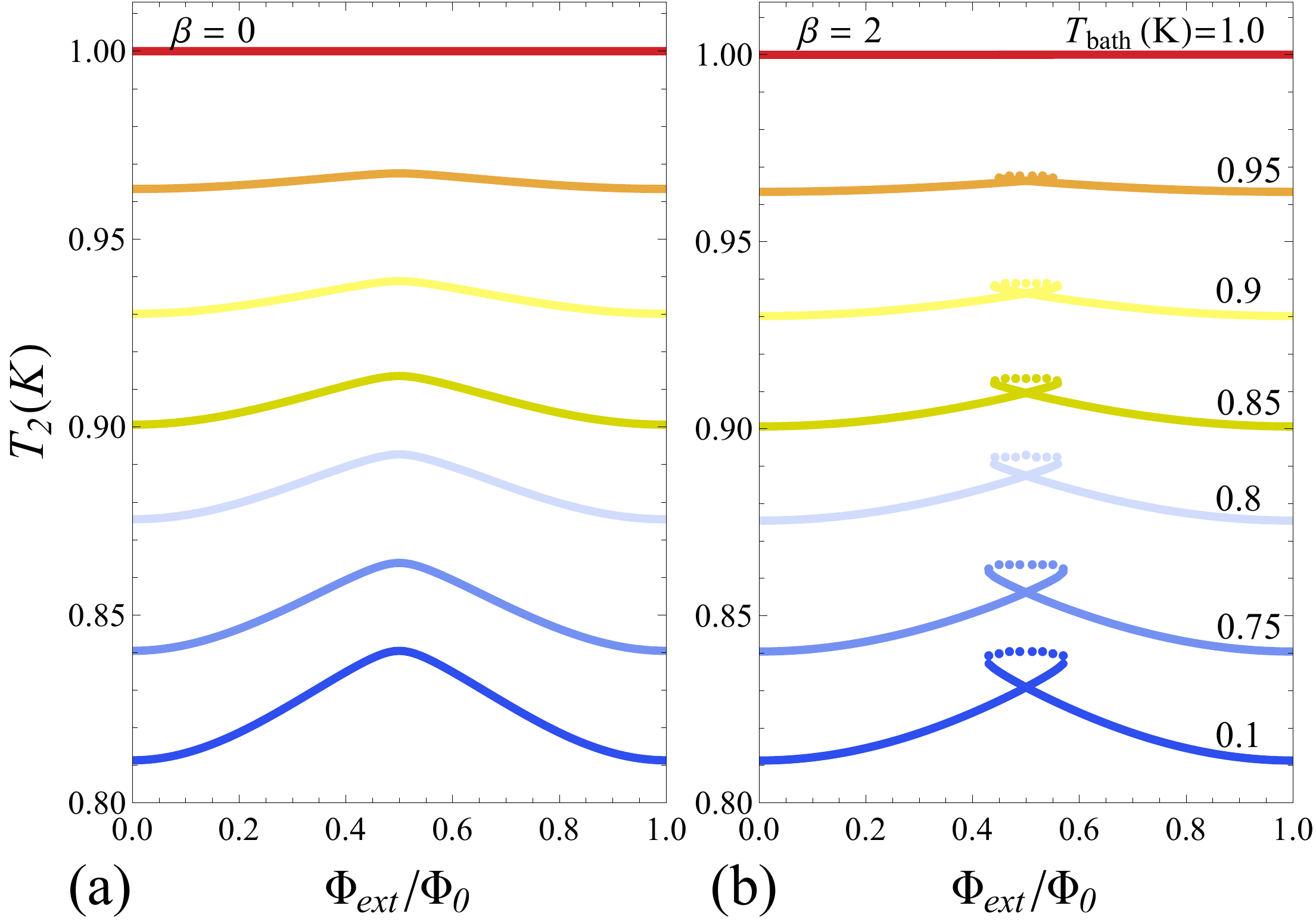}
\caption{(Color online) Quasiparticle temperature $T_2\;vs\;\Phi_{ext}$ calculated for a few values of $T_{bath}$ at $T_1 = 1 \text{K}$ for $\alpha=0.75$, $\beta=0.0$ and $\beta=2$, panel (a) and (b), respectively. Dotted curves represent the unstable states in the hysteretic mode.}
\label{Fig08}
\end{figure}

The role of the bath temperature is shown in Fig.~\ref{Fig08}, where $T_2(\Phi_{ext})$ is calculated for increasing $T_{bath}$ at $T_1=1K$, for $\beta=0.0$ and $\beta=2$, see panels (a) and (b), respectively. By increasing $T_{bath}$, the modulation of $T_2$ reduces, and vanishes for $T_{bath}=T_1$, see Fig.~\ref{Fig08}a. Accordingly, $\delta T_2$ decreases for $T_{bath}\to T_1$, since $\dot{Q}_{e-ph,2}$ enhances, the temperature drop reduces and the temperature-dependent energy gap in $S_2$ is suppressed. The hysteresis in $T_2(\Phi_{ext})$ is displayed in Fig.~\ref{Fig08}b for $\beta=2$.

We observe that, at the temperatures we are considering, proper values for the system parameters have to be chosen to avoid the degradation of the SQUID sensitivity due to thermal fluctuations~\cite{Cla04}. In fact, since the Josephson coupling energy $E_J$ should be much larger than the thermal energy $k_B T$, a lower limit for the critical current exists, namely $I_J/5\gtrsim2\pi k_b T/\Phi_0$~\cite{Cla88,Cla04}. Moreover, Nyquist noise imposes an upper limit on the SQUID inductance, such that $5L\lesssim\Phi_0^2/(4\pi^2 k_bT)$~\cite{Cla88,Cla04}. These constraints imply, for instance, $I_J\gtrsim0.2\mu \textup{A}$ and $L\lesssim1.5\textup{nH}$ at $T=1\textup{K}$. Obviously, the value of the inductance depends on the applications in which the SQUID is employed. In fact, although a large area of the ring, corresponding to a large value of $L$, may deteriorate the SQUID performance, it is advantageous to increase the sensitivity of a SQUID-based magnetic flux detector, since small field variations give large flux changes. 

Conversely, for memory applications, robustness against small external field fluctuations is desirable. In fact, a system showing a clear hysteresis can promptly find applications as memory elements. In superconducting devices it is natural to use persistent currents or magnetic flux in a superconducting loop for binary information storage~\cite{Mat80}. Specifically, in a SQUID the logical ``0'' and ``1'' usually correspond to zero and a single flux quantum in the loop, respectively. More recently, other superconducting tunnel junction-based memory elements were suggested~\cite{Peo14,She16,Sal16,GuaSol16,Pfe16}. A memory based on a thermally-biased inductive SQUID could take advantage of the clear hysteretic behavior of the temperature of the cold electrode for proper values of the external flux. For instance, the thermal jump $\Delta T_2\sim20\textup{mK}$, for $T_1=0.84\textup{K}$, $T_{bath}=0.1\textup{K}$, and $\Phi_{ext}\simeq0.63\Phi_0$, shown in Fig.~\ref{Fig07}b allows to clearly distinguish a ``heat-bit 1'', associated to the state with $k=0$, from a ``heat-bit 0'', for the state with $k=1$.

It is worth noting that the dynamics of a SQUID in the so-called adiabatic regime strongly depends on whether the frequency $\omega_{ext}$ of the external magnetic flux is smaller than both the cut-off frequency of the SQUID loop, $\omega_{cut}=R/L$, and the junction characteristic frequency~\cite{Bar82}, $\omega_c=2\pi R_aI^a_J/\Phi_0$. The time of a quantum transition, $k\to k\pm1$, as the flux through the SQUID suddenly changes, is given by $\omega^{-1}_{cut}$ or by $\omega^{-1}_{c}$, depending on which is larger~\cite{Cla04}. Therefore, in the adiabatic regime the sweep frequency $\omega_{ext}$ must be much slower than the characteristic time for a flux transition.

Finally, the speed of modulation of the temperature $T_2$ mainly depends on the relaxation time $\tau_{eph}$ required by the quasiparticle in $S_2$ to thermalize with lattice phonons, since the $R_a C$ time constant of the junctions forming the SQUID can be reduced more than $\tau_{eph}$ by properly choosing the system parameters. In particular, in the $0.5\textup{K} \div 1\textup{K}$ temperature range, $\tau_{eph}^{-1}$ is of the order of $\sim 1\textup{MHz}$ to $10\textup{MHz}$ for Al~\cite{GiaMar12,Kap76}, whereas at lower $T_{bath}$ it is drastically reduced owing to increased electron-phonon relaxation time~\cite{Kap76,Bar08}. However, we stress that $\tau_{eph}^{-1}$ can be enhanced by using other superconductors with higher electron-phonon coupling than Al, like, for instance, Tantalum~\cite{GiaMar12,Bar08}. Moreover, the use of superconductors with higher $T_c$, allows higher working temperatures, resulting in a further enhancement of the electron-phonon relaxation frequency. Finally, the fine-tuning of the system could allow memory applications also at GHz frequencies (see Appendix~\ref{AppA}).

\section{Conclusions}
\label{Conclusions}

We have studied the modulation of the temperature in a thermally-biased SQUID with a non-negligible inductance of the superconducting ring, when the external magnetic flux through the device, $\Phi_{ext}$, is changed. Specifically, we analysed the thermal current flowing in the SQUID by varying the values of the hysteresis parameter $\beta$, which is proportional to the product of the inductance and the critical current of a JJ, and the ratio $\alpha$ between the JJs critical currents. Moreover, we investigated the steady temperature, $T_2$, of the cold electrode for several temperatures of the heater and the thermal bath. For proper values of $\beta$, the SQUID behaves hysteretically as the external flux is properly swept. We observe temperature modulation as a function of $\Phi_{ext}$ and hysteretic transitions in the thermal current flowing through the junctions. This hysteretic behavior directly reflects on the temperature $T_2$, as the thermal contact with both the other electrode, i.e., the heater, and a phonon bath are taken into account in the thermal model. 

Accordingly, as $\Phi_{ext}$ induces a transition in the number of flux quanta through the SQUID ring, pronounced jumps in $T_2$ occur, up to $\Delta T_2\sim20\textup{mK}$, for $T_1=0.84\textup{K}$ and $T_{bath}=0.1\textup{K}$ in a realistic Al-based proposed setup. The emergence of this thermal hysteresis suggests the use of a thermally-biased inductive SQUID as a memory element, in which the input/output related variables are the external magnetic flux and the temperature of a branch of the SQUID. Such memory could work even in a range of frequencies on the order of GHz by properly choosing the superconductors forming the SQUID.
The proposed systems could be easily implemented by standard nanofabrication techniques through the setup proposed for the SQUID-based Josephson heat interferometer~\cite{Gia12}.

\section{Acknowledgments}
\label{acknowledgments}

C.G. and P.S. have received funding from the European Union FP7/2007-2013 under REA
Grant agreement No. 630925 -- COHEAT and from MIUR-FIRB2013 -- Project Coca (Grant
No.~RBFR1379UX). 
F.G. acknowledges the European Research Council under the European Union's Seventh Framework Program (FP7/2007-2013)/ERC Grant agreement No.~615187-COMANCHE for partial financial support. M.D. acknowledges support from Department of Energy under Grant No.~DE-FG02-05ER46204.

\begin{figure}[t!!]
\centering
\includegraphics[width=\columnwidth]{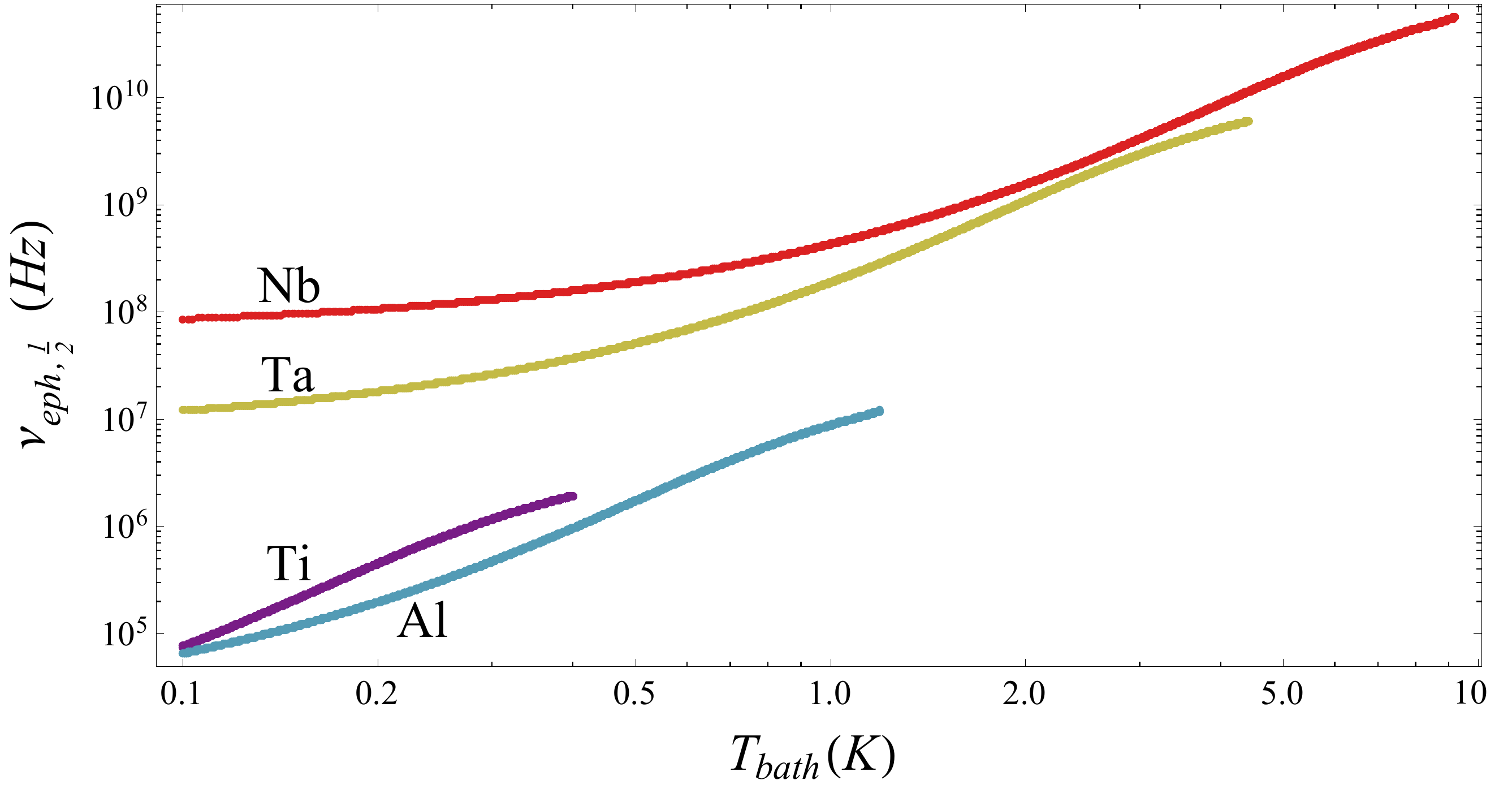}
\caption{(Color online) Electron-phonon relaxation frequency, $\nu_{eph,1/2}=\tau^{-1}_{eph,1/2}$, as a function of the bath temperature, $T_{bath}$, for $N_F=10^{47}\textup{J}^{-1}\textup{m}^{-3}$ and by choosing $T_{hot}=(T_{bath}+T_c)/2$, for several superconductors, i.e., Ti, Al, Ta, and Nb.}
\label{Fig09}
\end{figure}

\appendix
\section{Heat relaxation time scale}
\label{AppA}
The thermal relaxation time scale of the system can be estimated by the time for the superconductor, at the temperature $T_{hot}$, to reach the temperature $(T_{bath}+T_{hot})/2$, as the thermal contact with a phonon bath is taken into account. The thermal balance equation for a superconductor in thermal contact with a phonon bath can be written as
\begin{equation}
-\dot{Q}_{e-ph}\left ( T,T_{bath}\right )=C_v(T)\frac{\mathrm{d} T}{\mathrm{d} t},
\label{ThermalBalanceEq}
\end{equation}
where $\dot{Q}_{e-ph}$ is given by Eq.~(\ref{Qe-ph}), and the rhs fo Eq.~(\ref{ThermalBalanceEq}) represents the variations of the internal energy of the system.

In Eq.~(\ref{ThermalBalanceEq}), $C_v(T)$ is the heat capacity given by~\cite{Kus04}
\begin{equation}
C_v(T)=\mathcal{V}T\frac{\mathrm{d} S(T)}{\mathrm{d} T}
\label{HeatCapacity}
\end{equation}
where $\mathcal{V}$ is the volume and $S(T)$ is the electronic entropy of the superconductor~\cite{Kus04,Rab08,Sol16}
\begin{eqnarray}\nonumber
S=&-&4k_BN_F\int_{0}^{\infty}d\varepsilon \mathcal{N}(\varepsilon,T) \{ \left [ 1-f(\varepsilon,T) \right ] \log\left [ 1-f(\varepsilon,T) \right ]\\
&+&f(\varepsilon,T) \log f(\varepsilon,T) \},
\label{Entropy}
\end{eqnarray}

$N_F$ being the density of states at the Fermi energy.
From Eq.~(\ref{ThermalBalanceEq}), the electron-phonon relaxation time from the temperature $T_{hot}$ to the temperature $(T_{bath}+T_{hot})/2$ can be evaluated as
\begin{equation}
\tau_{eph,1/2}=\int_{\frac{T_{bath}+T_{hot}}{2}}^{T_{hot}}\frac{C_v(T)}{\dot{Q}_{e-ph}\left ( T,T_{bath}\right )}dT.
\label{taueph}
\end{equation}

From Eqs.~(\ref{Qe-ph}) and (\ref{taueph}), we observe that the electron-phonon relaxation frequency, $\nu_{eph,1/2}=\tau^{-1}_{eph,1/2}$, depends on both the working temperatures and the characteristics of the superconductor, such as the electron-phonon coupling constant $\Sigma$ and the critical temperature $T_c$. \\
\indent The behaviour of $\nu_{eph,1/2}$ as a function of the bath temperature, for $N_F=10^{47}\textup{J}^{-1}\textup{m}^{-3}$ and by choosing $T_{hot}=(T_{bath}+T_c)/2$, is shown in Fig.~\ref{Fig09} for several superconductors. Specifically, results in Fig.~\ref{Fig09} are obtained for Ti ($T_c=0.4\text{K}$, $\Sigma=1.33\:10^9\textup{Wm}^{-3}\textup{K}^{-5}$~\cite{Gia06}), Al ($T_c=1.19\text{K}$, $\Sigma=0.3\:10^9\textup{Wm}^{-3}\textup{K}^{-5}$~\cite{Gia06}), Ta ($T_c=4.43\text{K}$, $\Sigma=3\:10^9\textup{Wm}^{-3}\textup{K}^{-5}$~\cite{GiaMar12,Bar08}), and Nb ($T_c=9.2\text{K}$, by supposing $\Sigma=3\:10^9\textup{Wm}^{-3}\textup{K}^{-5}$). We observe that $\nu_{eph,1/2}\lesssim 10\text{MHz}$ for Al, whereas e-ph relaxation frequencies of the order of GHz can be achieved at $T_{bath}\sim2K$ for both Ta and Nb.


\begin{thebibliography}{34}%
\makeatletter
\providecommand \@ifxundefined [1]{%
 \@ifx{#1\undefined}
}%
\providecommand \@ifnum [1]{%
 \ifnum #1\expandafter \@firstoftwo
 \else \expandafter \@secondoftwo
 \fi
}%
\providecommand \@ifx [1]{%
 \ifx #1\expandafter \@firstoftwo
 \else \expandafter \@secondoftwo
 \fi
}%
\providecommand \natexlab [1]{#1}%
\providecommand \enquote [1]{``#1''}%
\providecommand \bibnamefont [1]{#1}%
\providecommand \bibfnamefont [1]{#1}%
\providecommand \citenamefont [1]{#1}%
\providecommand \href@noop [0]{\@secondoftwo}%
\providecommand \href [0]{\begingroup \@sanitize@url \@href}%
\providecommand \@href[1]{\@@startlink{#1}\@@href}%
\providecommand \@@href[1]{\endgroup#1\@@endlink}%
\providecommand \@sanitize@url [0]{\catcode `\\12\catcode `\$12\catcode
 `\&12\catcode `\#12\catcode `\^12\catcode `\_12\catcode `\%12\relax}%
\providecommand \@@startlink[1]{}%
\providecommand \@@endlink[0]{}%
\providecommand \url [0]{\begingroup\@sanitize@url \@url }%
\providecommand \@url [1]{\endgroup\@href {#1}{\urlprefix }}%
\providecommand \urlprefix [0]{URL }%
\providecommand \Eprint [0]{\href }%
\providecommand \doibase [0]{http://dx.doi.org/}%
\providecommand \selectlanguage [0]{\@gobble}%
\providecommand \bibinfo [0]{\@secondoftwo}%
\providecommand \bibfield [0]{\@secondoftwo}%
\providecommand \translation [1]{[#1]}%
\providecommand \BibitemOpen [0]{}%
\providecommand \bibitemStop [0]{}%
\providecommand \bibitemNoStop [0]{.\EOS\space}%
\providecommand \EOS [0]{\spacefactor3000\relax}%
\providecommand \BibitemShut [1]{\csname bibitem#1\endcsname}%
\let\auto@bib@innerbib\@empty
\bibitem [{\citenamefont {Maki}\ and\ \citenamefont {Griffin}(1965)}]{Mak65}%
 \BibitemOpen
 \bibfield {author} {\bibinfo {author} {\bibfnamefont {K.}~\bibnamefont
 {Maki}}\ and\ \bibinfo {author} {\bibfnamefont {A.}~\bibnamefont {Griffin}},\
 }\href {\doibase 10.1103/PhysRevLett.15.921} {\bibfield {journal} {\bibinfo
 {journal} {Phys. Rev. Lett.}\ }\textbf {\bibinfo {volume} {15}},\ \bibinfo
 {pages} {921} (\bibinfo {year} {1965})}\BibitemShut {NoStop}%
\bibitem [{\citenamefont {Giazotto}\ and\ \citenamefont
 {Mart{\'i}nez-P{\'e}rez}(2012{\natexlab{a}})}]{Gia12}%
 \BibitemOpen
 \bibfield {author} {\bibinfo {author} {\bibfnamefont {F.}~\bibnamefont
 {Giazotto}}\ and\ \bibinfo {author} {\bibfnamefont {M.~J.}\ \bibnamefont
 {Mart{\'i}nez-P{\'e}rez}},\ }\href@noop {} {\bibfield {journal} {\bibinfo
 {journal} {Nature}\ }\textbf {\bibinfo {volume} {492}},\ \bibinfo {pages}
 {401} (\bibinfo {year} {2012}{\natexlab{a}})}\BibitemShut {NoStop}%
\bibitem [{\citenamefont {Mart{\'i}nez-P{\'e}rez}\ and\ \citenamefont
 {Giazotto}(2013{\natexlab{a}})}]{Mar13}%
 \BibitemOpen
 \bibfield {author} {\bibinfo {author} {\bibfnamefont {M.~J.}\ \bibnamefont
 {Mart{\'i}nez-P{\'e}rez}}\ and\ \bibinfo {author} {\bibfnamefont
 {F.}~\bibnamefont {Giazotto}},\ }\href {\doibase 10.1063/1.4794412}
 {\bibfield {journal} {\bibinfo {journal} {Appl. Phys. Lett.}\ }\textbf
 {\bibinfo {volume} {102}},\ \bibinfo {pages} {092602} (\bibinfo {year}
 {2013}{\natexlab{a}})}\BibitemShut {NoStop}%
\bibitem [{\citenamefont {Mart{\'i}nez-P{\'e}rez}\ and\ \citenamefont
 {Giazotto}(2013{\natexlab{b}})}]{MarGia13}%
 \BibitemOpen
 \bibfield {author} {\bibinfo {author} {\bibfnamefont {M.~J.}\ \bibnamefont
 {Mart{\'i}nez-P{\'e}rez}}\ and\ \bibinfo {author} {\bibfnamefont
 {F.}~\bibnamefont {Giazotto}},\ }\href {\doibase 10.1063/1.4804550}
 {\bibfield {journal} {\bibinfo {journal} {Appl. Phys. Lett.}\ }\textbf
 {\bibinfo {volume} {102}},\ \bibinfo {pages} {182602} (\bibinfo {year}
 {2013}{\natexlab{b}})}\BibitemShut {NoStop}%
\bibitem [{\citenamefont {Mart{\'i}nez-P{\'e}rez}\ and\ \citenamefont
 {Giazotto}(2014)}]{Mar14}%
 \BibitemOpen
 \bibfield {author} {\bibinfo {author} {\bibfnamefont {M.~J.}\ \bibnamefont
 {Mart{\'i}nez-P{\'e}rez}}\ and\ \bibinfo {author} {\bibfnamefont
 {F.}~\bibnamefont {Giazotto}},\ }\href@noop {} {\bibfield {journal}
 {\bibinfo {journal} {Nat. Commun.}\ }\textbf {\bibinfo {volume} {5}},\
 \bibinfo {pages} {3579} (\bibinfo {year} {2014})}\BibitemShut {NoStop}%
\bibitem [{\citenamefont {Mart{\'i}nez-P{\'e}rez}\ \emph
 {et~al.}(2014)\citenamefont {Mart{\'i}nez-P{\'e}rez}, \citenamefont
 {Solinas},\ and\ \citenamefont {Giazotto}}]{MarSol14}%
 \BibitemOpen
 \bibfield {author} {\bibinfo {author} {\bibfnamefont {M.~J.}\ \bibnamefont
 {Mart{\'i}nez-P{\'e}rez}}, \bibinfo {author} {\bibfnamefont {P.}~\bibnamefont
 {Solinas}}, \ and\ \bibinfo {author} {\bibfnamefont {F.}~\bibnamefont
 {Giazotto}},\ }\href {\doibase 10.1007/s10909-014-1132-6} {\bibfield
 {journal} {\bibinfo {journal} {J. Low Temp. Phys.}\ }\textbf {\bibinfo
 {volume} {175}},\ \bibinfo {pages} {813} (\bibinfo {year}
 {2014})}\BibitemShut {NoStop}%
\bibitem [{\citenamefont {Fornieri}\ \emph
 {et~al.}(2016{\natexlab{a}})\citenamefont {Fornieri}, \citenamefont {Blanc},
 \citenamefont {Bosisio}, \citenamefont {D'Ambrosio},\ and\ \citenamefont
 {Giazotto}}]{For16}%
 \BibitemOpen
 \bibfield {author} {\bibinfo {author} {\bibfnamefont {A.}~\bibnamefont
 {Fornieri}}, \bibinfo {author} {\bibfnamefont {C.}~\bibnamefont {Blanc}},
 \bibinfo {author} {\bibfnamefont {R.}~\bibnamefont {Bosisio}}, \bibinfo
 {author} {\bibfnamefont {S.}~\bibnamefont {D'Ambrosio}}, \ and\ \bibinfo
 {author} {\bibfnamefont {F.}~\bibnamefont {Giazotto}},\ }\href {\doibase
 10.1038/nnano.2015.281} {\bibfield {journal} {\bibinfo {journal} {Nat.
 Nanotechnol.}\ }\textbf {\bibinfo {volume} {11}},\ \bibinfo {pages} {258}
 (\bibinfo {year} {2016}{\natexlab{a}})}\BibitemShut {NoStop}%
\bibitem [{\citenamefont {Fornieri}\ \emph
 {et~al.}(2016{\natexlab{b}})\citenamefont {Fornieri}, \citenamefont
 {Timossi}, \citenamefont {Virtanen}, \citenamefont {Solinas},\ and\
 \citenamefont {Giazotto}}]{ForTim16}%
 \BibitemOpen
 \bibfield {author} {\bibinfo {author} {\bibfnamefont {A.}~\bibnamefont
 {Fornieri}}, \bibinfo {author} {\bibfnamefont {G.}~\bibnamefont {Timossi}},
 \bibinfo {author} {\bibfnamefont {P.}~\bibnamefont {Virtanen}}, \bibinfo
 {author} {\bibfnamefont {P.}~\bibnamefont {Solinas}}, \ and\ \bibinfo
 {author} {\bibfnamefont {F.}~\bibnamefont {Giazotto}},\ }\href@noop {}
 {\bibfield {journal} {\bibinfo {journal} {arXiv preprint arXiv:1607.02428}\
 } (\bibinfo {year} {2016}{\natexlab{b}})}\BibitemShut {NoStop}%
\bibitem [{\citenamefont {Fornieri}\ and\ \citenamefont
 {Giazotto}(2016)}]{ForGia16}%
 \BibitemOpen
 \bibfield {author} {\bibinfo {author} {\bibfnamefont {A.}~\bibnamefont
 {Fornieri}}\ and\ \bibinfo {author} {\bibfnamefont {F.}~\bibnamefont
 {Giazotto}},\ }\href@noop {} {\bibfield {journal} {\bibinfo {journal}
 {arXiv preprint arXiv:1610.01013}\ } (\bibinfo {year} {2016})}\BibitemShut
 {NoStop}%
\bibitem [{\citenamefont {Giazotto}\ and\ \citenamefont
 {Mart{\'i}nez-P{\'e}rez}(2012{\natexlab{b}})}]{GiaMar12}%
 \BibitemOpen
 \bibfield {author} {\bibinfo {author} {\bibfnamefont {F.}~\bibnamefont
 {Giazotto}}\ and\ \bibinfo {author} {\bibfnamefont {M.~J.}\ \bibnamefont
 {Mart{\'i}nez-P{\'e}rez}},\ }\href
 {http://scitation.aip.org/content/aip/journal/apl/101/10/10.1063/1.4750068}
 {\bibfield {journal} {\bibinfo {journal} {Appl. Phys. Lett.}\ }\textbf
 {\bibinfo {volume} {101}},\ \bibinfo {eid} {102601} (\bibinfo {year}
 {2012}{\natexlab{b}})}\BibitemShut {NoStop}%
\bibitem [{\citenamefont {Guttman}\ \emph {et~al.}(1997)\citenamefont
 {Guttman}, \citenamefont {Nathanson}, \citenamefont {Ben-Jacob},\ and\
 \citenamefont {Bergman}}]{Gut97}%
 \BibitemOpen
 \bibfield {author} {\bibinfo {author} {\bibfnamefont {G.~D.}\ \bibnamefont
 {Guttman}}, \bibinfo {author} {\bibfnamefont {B.}~\bibnamefont {Nathanson}},
 \bibinfo {author} {\bibfnamefont {E.}~\bibnamefont {Ben-Jacob}}, \ and\
 \bibinfo {author} {\bibfnamefont {D.~J.}\ \bibnamefont {Bergman}},\ }\href
 {\doibase 10.1103/PhysRevB.55.3849} {\bibfield {journal} {\bibinfo
 {journal} {Phys. Rev. B}\ }\textbf {\bibinfo {volume} {55}},\ \bibinfo
 {pages} {3849} (\bibinfo {year} {1997})}\BibitemShut {NoStop}%
\bibitem [{\citenamefont {Guttman}\ \emph {et~al.}(1998)\citenamefont
 {Guttman}, \citenamefont {Ben-Jacob},\ and\ \citenamefont {Bergman}}]{Gut98}%
 \BibitemOpen
 \bibfield {author} {\bibinfo {author} {\bibfnamefont {G.~D.}\ \bibnamefont
 {Guttman}}, \bibinfo {author} {\bibfnamefont {E.}~\bibnamefont {Ben-Jacob}},
 \ and\ \bibinfo {author} {\bibfnamefont {D.~J.}\ \bibnamefont {Bergman}},\
 }\href {\doibase 10.1103/PhysRevB.57.2717} {\bibfield {journal} {\bibinfo
 {journal} {Phys. Rev. B}\ }\textbf {\bibinfo {volume} {57}},\ \bibinfo
 {pages} {2717} (\bibinfo {year} {1998})}\BibitemShut {NoStop}%
\bibitem [{\citenamefont {Zhao}\ \emph {et~al.}(2003)\citenamefont {Zhao},
 \citenamefont {L\"ofwander},\ and\ \citenamefont {Sauls}}]{Zha03}%
 \BibitemOpen
 \bibfield {author} {\bibinfo {author} {\bibfnamefont {E.}~\bibnamefont
 {Zhao}}, \bibinfo {author} {\bibfnamefont {T.}~\bibnamefont {L\"ofwander}}, \
 and\ \bibinfo {author} {\bibfnamefont {J.~A.}\ \bibnamefont {Sauls}},\ }\href
 {\doibase 10.1103/PhysRevLett.91.077003} {\bibfield {journal} {\bibinfo
 {journal} {Phys. Rev. Lett.}\ }\textbf {\bibinfo {volume} {91}},\ \bibinfo
 {pages} {077003} (\bibinfo {year} {2003})}\BibitemShut {NoStop}%
\bibitem [{\citenamefont {Zhao}\ \emph {et~al.}(2004)\citenamefont {Zhao},
 \citenamefont {L\"ofwander},\ and\ \citenamefont {Sauls}}]{Zha04}%
 \BibitemOpen
 \bibfield {author} {\bibinfo {author} {\bibfnamefont {E.}~\bibnamefont
 {Zhao}}, \bibinfo {author} {\bibfnamefont {T.}~\bibnamefont {L\"ofwander}}, \
 and\ \bibinfo {author} {\bibfnamefont {J.~A.}\ \bibnamefont {Sauls}},\ }\href
 {\doibase 10.1103/PhysRevB.69.134503} {\bibfield {journal} {\bibinfo
 {journal} {Phys. Rev. B}\ }\textbf {\bibinfo {volume} {69}},\ \bibinfo
 {pages} {134503} (\bibinfo {year} {2004})}\BibitemShut {NoStop}%
\bibitem [{\citenamefont {Golubev}\ \emph {et~al.}(2013)\citenamefont
 {Golubev}, \citenamefont {Faivre},\ and\ \citenamefont {Pekola}}]{Gol13}%
 \BibitemOpen
 \bibfield {author} {\bibinfo {author} {\bibfnamefont {D.}~\bibnamefont
 {Golubev}}, \bibinfo {author} {\bibfnamefont {T.}~\bibnamefont {Faivre}}, \
 and\ \bibinfo {author} {\bibfnamefont {J.~P.}\ \bibnamefont {Pekola}},\
 }\href {\doibase 10.1103/PhysRevB.87.094522} {\bibfield {journal} {\bibinfo
 {journal} {Phys. Rev. B}\ }\textbf {\bibinfo {volume} {87}},\ \bibinfo
 {pages} {094522} (\bibinfo {year} {2013})}\BibitemShut {NoStop}%
\bibitem [{\citenamefont {Giazotto}\ \emph {et~al.}(2006)\citenamefont
 {Giazotto}, \citenamefont {Heikkil\"a}, \citenamefont {Luukanen},
 \citenamefont {Savin},\ and\ \citenamefont {Pekola}}]{Gia06}%
 \BibitemOpen
 \bibfield {author} {\bibinfo {author} {\bibfnamefont {F.}~\bibnamefont
 {Giazotto}}, \bibinfo {author} {\bibfnamefont {T.~T.}\ \bibnamefont
 {Heikkil\"a}}, \bibinfo {author} {\bibfnamefont {A.}~\bibnamefont
 {Luukanen}}, \bibinfo {author} {\bibfnamefont {A.~M.}\ \bibnamefont {Savin}},
 \ and\ \bibinfo {author} {\bibfnamefont {J.~P.}\ \bibnamefont {Pekola}},\
 }\href {\doibase 10.1103/RevModPhys.78.217} {\bibfield {journal} {\bibinfo
 {journal} {Rev. Mod. Phys.}\ }\textbf {\bibinfo {volume} {78}},\ \bibinfo
 {pages} {217} (\bibinfo {year} {2006})}\BibitemShut {NoStop}%
\bibitem [{\citenamefont {Frank}\ and\ \citenamefont {Krech}(1997)}]{Fra97}%
 \BibitemOpen
 \bibfield {author} {\bibinfo {author} {\bibfnamefont {B.}~\bibnamefont
 {Frank}}\ and\ \bibinfo {author} {\bibfnamefont {W.}~\bibnamefont {Krech}},\
 }\href {\doibase http://dx.doi.org/10.1016/S0375-9601(97)00627-0} {\bibfield
 {journal} {\bibinfo {journal} {Phys. Lett. A}\ }\textbf {\bibinfo {volume}
 {235}},\ \bibinfo {pages} {281 } (\bibinfo {year} {1997})}\BibitemShut
 {NoStop}%
\bibitem [{\citenamefont {Barone}\ and\ \citenamefont
 {Patern\`{o}}(1982)}]{Bar82}%
 \BibitemOpen
 \bibfield {author} {\bibinfo {author} {\bibfnamefont {A.}~\bibnamefont
 {Barone}}\ and\ \bibinfo {author} {\bibfnamefont {G.}~\bibnamefont
 {Patern\`{o}}},\ }\href@noop {} {\emph {\bibinfo {title} {Physics and
 Applications of the Josephson Effect}}}\ (\bibinfo {publisher} {Wiley, New
 York},\ \bibinfo {year} {1982})\BibitemShut {NoStop}%
 \bibitem [{\citenamefont {Dynes}\ \emph {et~al.}(1978)\citenamefont {Dynes},
 \citenamefont {Narayanamurti},\ and\ \citenamefont {Garno}}]{Dyn78}%
 \BibitemOpen
 \bibfield {author} {\bibinfo {author} {\bibfnamefont {R.~C.}\ \bibnamefont
 {Dynes}}, \bibinfo {author} {\bibfnamefont {V.}~\bibnamefont
 {Narayanamurti}}, \ and\ \bibinfo {author} {\bibfnamefont {J.~P.}\
 \bibnamefont {Garno}},\ }\href {\doibase 10.1103/PhysRevLett.41.1509}
 {\bibfield {journal} {\bibinfo {journal} {Phys. Rev. Lett.}\ }\textbf
 {\bibinfo {volume} {41}},\ \bibinfo {pages} {1509} (\bibinfo {year}
 {1978})}\BibitemShut {NoStop}%
 \bibitem [{\citenamefont {Mart{\'i}nez-P{\'e}rez}\ \emph
 {et~al.}(2015)\citenamefont {Mart{\'i}nez-P{\'e}rez}, \citenamefont
 {Fornieri},\ and\ \citenamefont {Giazotto}}]{Mar15}%
 \BibitemOpen
 \bibfield {author} {\bibinfo {author} {\bibfnamefont {M.~J.}\ \bibnamefont
 {Mart{\'i}nez-P{\'e}rez}}, \bibinfo {author} {\bibfnamefont {A.}~\bibnamefont
 {Fornieri}}, \ and\ \bibinfo {author} {\bibfnamefont {F.}~\bibnamefont
 {Giazotto}},\ }\href@noop {} {\bibfield {journal} {\bibinfo {journal}
 {Nat. Nanotechnol.}\ }\textbf {\bibinfo {volume} {10}},\ \bibinfo {pages}
 {303} (\bibinfo {year} {2015})}\BibitemShut {NoStop}%
\bibitem [{\citenamefont {Fornieri}\ \emph
 {et~al.}(2016{\natexlab{c}})\citenamefont {Fornieri}, \citenamefont
 {Timossi}, \citenamefont {Bosisio}, \citenamefont {Solinas},\ and\
 \citenamefont {Giazotto}}]{ForTimBos16}%
 \BibitemOpen
 \bibfield {author} {\bibinfo {author} {\bibfnamefont {A.}~\bibnamefont
 {Fornieri}}, \bibinfo {author} {\bibfnamefont {G.}~\bibnamefont {Timossi}},
 \bibinfo {author} {\bibfnamefont {R.}~\bibnamefont {Bosisio}}, \bibinfo
 {author} {\bibfnamefont {P.}~\bibnamefont {Solinas}}, \ and\ \bibinfo
 {author} {\bibfnamefont {F.}~\bibnamefont {Giazotto}},\ }\href {\doibase
 10.1103/PhysRevB.93.134508} {\bibfield {journal} {\bibinfo {journal} {Phys.
 Rev. B}\ }\textbf {\bibinfo {volume} {93}},\ \bibinfo {pages} {134508}
 (\bibinfo {year} {2016}{\natexlab{c}})}\BibitemShut {NoStop}%
\bibitem [{\citenamefont {Giazotto}\ and\ \citenamefont
 {Pekola}(2005)}]{Gia05}%
 \BibitemOpen
 \bibfield {author} {\bibinfo {author} {\bibfnamefont {F.}~\bibnamefont
 {Giazotto}}\ and\ \bibinfo {author} {\bibfnamefont {J.~P.}\ \bibnamefont
 {Pekola}},\ }\href
 {http://scitation.aip.org/content/aip/journal/jap/97/2/10.1063/1.1833576}
 {\bibfield {journal} {\bibinfo {journal} {J. Appl. Phys.}\ }\textbf
 {\bibinfo {volume} {97}},\ \bibinfo {eid} {023908} (\bibinfo {year}
 {2005})}\BibitemShut {NoStop}%
\bibitem [{\citenamefont {Tirelli}\ \emph {et~al.}(2008)\citenamefont
 {Tirelli}, \citenamefont {Savin}, \citenamefont {Garcia}, \citenamefont
 {Pekola}, \citenamefont {Beltram},\ and\ \citenamefont {Giazotto}}]{Tir08}%
 \BibitemOpen
 \bibfield {author} {\bibinfo {author} {\bibfnamefont {S.}~\bibnamefont
 {Tirelli}}, \bibinfo {author} {\bibfnamefont {A.~M.}\ \bibnamefont {Savin}},
 \bibinfo {author} {\bibfnamefont {C.~P.}\ \bibnamefont {Garcia}}, \bibinfo
 {author} {\bibfnamefont {J.~P.}\ \bibnamefont {Pekola}}, \bibinfo {author}
 {\bibfnamefont {F.}~\bibnamefont {Beltram}}, \ and\ \bibinfo {author}
 {\bibfnamefont {F.}~\bibnamefont {Giazotto}},\ }\href {\doibase
 10.1103/PhysRevLett.101.077004} {\bibfield {journal} {\bibinfo {journal}
 {Phys. Rev. Lett.}\ }\textbf {\bibinfo {volume} {101}},\ \bibinfo {pages}
 {077004} (\bibinfo {year} {2008})}\BibitemShut {NoStop}%
\bibitem [{\citenamefont {Bosisio}\ \emph {et~al.}(2016)\citenamefont
 {Bosisio}, \citenamefont {Solinas}, \citenamefont {Braggio},\ and\
 \citenamefont {Giazotto}}]{Bos16}%
 \BibitemOpen
 \bibfield {author} {\bibinfo {author} {\bibfnamefont {R.}~\bibnamefont
 {Bosisio}}, \bibinfo {author} {\bibfnamefont {P.}~\bibnamefont {Solinas}},
 \bibinfo {author} {\bibfnamefont {A.}~\bibnamefont {Braggio}}, \ and\
 \bibinfo {author} {\bibfnamefont {F.}~\bibnamefont {Giazotto}},\ }\href
 {\doibase 10.1103/PhysRevB.93.144512} {\bibfield {journal} {\bibinfo
 {journal} {Phys. Rev. B}\ }\textbf {\bibinfo {volume} {93}},\ \bibinfo
 {pages} {144512} (\bibinfo {year} {2016})}\BibitemShut {NoStop}%
\bibitem [{\citenamefont {Majer}\ \emph {et~al.}(2002)\citenamefont {Majer},
 \citenamefont {Butcher},\ and\ \citenamefont {Mooij}}]{Maj02}%
 \BibitemOpen
 \bibfield {author} {\bibinfo {author} {\bibfnamefont {J.~B.}\ \bibnamefont
 {Majer}}, \bibinfo {author} {\bibfnamefont {J.~R.}\ \bibnamefont {Butcher}},
 \ and\ \bibinfo {author} {\bibfnamefont {J.~E.}\ \bibnamefont {Mooij}},\
 }\href {\doibase http://dx.doi.org/10.1063/1.1478150} {\bibfield {journal}
 {\bibinfo {journal} {Appl. Phys. Lett.}\ }\textbf {\bibinfo {volume} {80}},\
 \bibinfo {pages} {3638} (\bibinfo {year} {2002})}\BibitemShut {NoStop}%
\bibitem [{\citenamefont {Clarke}\ and\ \citenamefont
 {Braginski}(2004)}]{Cla04}%
 \BibitemOpen
 \bibfield {author} {\bibinfo {author} {\bibfnamefont {J.}~\bibnamefont
 {Clarke}}\ and\ \bibinfo {author} {\bibfnamefont {A.}~\bibnamefont
 {Braginski}},\ }\href@noop {} {\emph {\bibinfo {title} {The SQUID Handbook:
 Fundamentals and Technology of SQUIDs and SQUID Systems}}},\ \bibinfo
 {series} {The SQUID Handbook}\ No.\ \bibinfo {number} {v. 1}\ (\bibinfo
 {publisher} {Wiley},\ \bibinfo {year} {2004})\BibitemShut {NoStop}%
\bibitem [{\citenamefont {Bo}\ \emph {et~al.}(2004)\citenamefont {Bo},
 \citenamefont {Zhong-Kui}, \citenamefont {Shu-Chao}, \citenamefont
 {Yuan-Dong},\ and\ \citenamefont {Fu-Ren}}]{Bo04}%
 \BibitemOpen
 \bibfield {author} {\bibinfo {author} {\bibfnamefont {M.}~\bibnamefont
 {Bo}}, \bibinfo {author} {\bibfnamefont {T.}~\bibnamefont {Zhong-Kui}},
 \bibinfo {author} {\bibfnamefont {M.}~\bibnamefont {Shu-Chao}}, \bibinfo
 {author} {\bibfnamefont {D.}~\bibnamefont {Yuan-Dong}}, \ and\ \bibinfo
 {author} {\bibfnamefont {W.}~\bibnamefont {Fu-Ren}},\ }\href
 {http://stacks.iop.org/1009-1963/13/i=8/a=008} {\bibfield {journal}
 {\bibinfo {journal} {Chin. Phys.}\ }\textbf {\bibinfo {volume} {13}},\
 \bibinfo {pages} {1226} (\bibinfo {year} {2004})}\BibitemShut {NoStop}%
\bibitem [{\citenamefont {Timofeev}\ \emph {et~al.}(2009)\citenamefont
 {Timofeev}, \citenamefont {Garc\'{\i}a}, \citenamefont {Kopnin},
 \citenamefont {Savin}, \citenamefont {Meschke}, \citenamefont {Giazotto},\
 and\ \citenamefont {Pekola}}]{Pek09}%
 \BibitemOpen
 \bibfield {author} {\bibinfo {author} {\bibfnamefont {A.~V.}\ \bibnamefont
 {Timofeev}}, \bibinfo {author} {\bibfnamefont {C.~P.}\ \bibnamefont
 {Garc\'{\i}a}}, \bibinfo {author} {\bibfnamefont {N.~B.}\ \bibnamefont
 {Kopnin}}, \bibinfo {author} {\bibfnamefont {A.~M.}\ \bibnamefont {Savin}},
 \bibinfo {author} {\bibfnamefont {M.}~\bibnamefont {Meschke}}, \bibinfo
 {author} {\bibfnamefont {F.}~\bibnamefont {Giazotto}}, \ and\ \bibinfo
 {author} {\bibfnamefont {J.~P.}\ \bibnamefont {Pekola}},\ }\href {\doibase
 10.1103/PhysRevLett.102.017003} {\bibfield {journal} {\bibinfo {journal}
 {Phys. Rev. Lett.}\ }\textbf {\bibinfo {volume} {102}},\ \bibinfo {pages}
 {017003} (\bibinfo {year} {2009})}\BibitemShut {NoStop}%
\bibitem [{\citenamefont {Clarke}\ and\ \citenamefont {Koch}(1988)}]{Cla88}%
 \BibitemOpen
 \bibfield {author} {\bibinfo {author} {\bibfnamefont {J.}~\bibnamefont
 {Clarke}}\ and\ \bibinfo {author} {\bibfnamefont {R.~H.}\ \bibnamefont
 {Koch}},\ }\href {\doibase 10.1126/science.242.4876.217} {\bibfield
 {journal} {\bibinfo {journal} {Science}\ }\textbf {\bibinfo {volume}
 {242}},\ \bibinfo {pages} {217} (\bibinfo {year} {1988})}\BibitemShut
 {NoStop}%
\bibitem [{\citenamefont {Matisoo}(1980)}]{Mat80}%
 \BibitemOpen
 \bibfield {author} {\bibinfo {author} {\bibfnamefont {J.}~\bibnamefont
 {Matisoo}},\ }\href {\doibase 10.1147/rd.242.0113} {\bibfield {journal}
 {\bibinfo {journal} {IBM J. Res. Dev.}\ }\textbf {\bibinfo {volume} {24}},\
 \bibinfo {pages} {113} (\bibinfo {year} {1980})}\BibitemShut {NoStop}%
\bibitem [{\citenamefont {Peotta}\ and\ \citenamefont
 {Di~Ventra}(2014)}]{Peo14}%
 \BibitemOpen
 \bibfield {author} {\bibinfo {author} {\bibfnamefont {S.}~\bibnamefont
 {Peotta}}\ and\ \bibinfo {author} {\bibfnamefont {M.}~\bibnamefont
 {Di~Ventra}},\ }\href {\doibase 10.1103/PhysRevApplied.2.034011} {\bibfield
 {journal} {\bibinfo {journal} {Phys. Rev. Applied}\ }\textbf {\bibinfo
 {volume} {2}},\ \bibinfo {pages} {034011} (\bibinfo {year}
 {2014})}\BibitemShut {NoStop}%
\bibitem [{\citenamefont {Shevchenko}\ \emph {et~al.}(2016)\citenamefont
 {Shevchenko}, \citenamefont {Pershin},\ and\ \citenamefont {Nori}}]{She16}%
 \BibitemOpen
 \bibfield {author} {\bibinfo {author} {\bibfnamefont {S.~N.}\ \bibnamefont
 {Shevchenko}}, \bibinfo {author} {\bibfnamefont {Y.~V.}\ \bibnamefont
 {Pershin}}, \ and\ \bibinfo {author} {\bibfnamefont {F.}~\bibnamefont
 {Nori}},\ }\href {\doibase 10.1103/PhysRevApplied.6.014006} {\bibfield
 {journal} {\bibinfo {journal} {Phys. Rev. Applied}\ }\textbf {\bibinfo
 {volume} {6}},\ \bibinfo {pages} {014006} (\bibinfo {year}
 {2016})}\BibitemShut {NoStop}%
\bibitem [{\citenamefont {Salmilehto}\ \emph {et~al.}(2016)\citenamefont
 {Salmilehto}, \citenamefont {Deppe}, \citenamefont {Di~Ventra}, \citenamefont
 {Sanz},\ and\ \citenamefont {Solano}}]{Sal16}%
 \BibitemOpen
 \bibfield {author} {\bibinfo {author} {\bibfnamefont {J.}~\bibnamefont
 {Salmilehto}}, \bibinfo {author} {\bibfnamefont {F.}~\bibnamefont {Deppe}},
 \bibinfo {author} {\bibfnamefont {M.}~\bibnamefont {Di~Ventra}}, \bibinfo
 {author} {\bibfnamefont {M.}~\bibnamefont {Sanz}}, \ and\ \bibinfo {author}
 {\bibfnamefont {E.}~\bibnamefont {Solano}},\ }\href@noop {} {\bibfield
 {journal} {\bibinfo {journal} {arXiv preprint arXiv:1603.04487}\ } (\bibinfo
 {year} {2016})}\BibitemShut {NoStop}%
\bibitem [{\citenamefont {Guarcello}\ \emph {et~al.}(2016)\citenamefont
 {Guarcello}, \citenamefont {Solinas}, \citenamefont {Di~Ventra},\ and\
 \citenamefont {Giazotto}}]{GuaSol16}%
 \BibitemOpen
 \bibfield {author} {\bibinfo {author} {\bibfnamefont {C.}~\bibnamefont
 {Guarcello}}, \bibinfo {author} {\bibfnamefont {P.}~\bibnamefont {Solinas}},
 \bibinfo {author} {\bibfnamefont {M.}~\bibnamefont {Di~Ventra}}, \ and\
 \bibinfo {author} {\bibfnamefont {F.}~\bibnamefont {Giazotto}},\ }\href@noop
 {} {\bibfield {journal} {\bibinfo {journal} {arXiv preprint
 arXiv:1610.06807}\ } (\bibinfo {year} {2016})}\BibitemShut {NoStop}%
\bibitem [{\citenamefont {Pfeiffer}\ \emph {et~al.}(2016)\citenamefont
 {Pfeiffer}, \citenamefont {Egusquiza}, \citenamefont {Di~Ventra},
 \citenamefont {Sanz},\ and\ \citenamefont {Solano}}]{Pfe16}%
 \BibitemOpen
 \bibfield {author} {\bibinfo {author} {\bibfnamefont {P.}~\bibnamefont
 {Pfeiffer}}, \bibinfo {author} {\bibfnamefont {I.}~\bibnamefont {Egusquiza}},
 \bibinfo {author} {\bibfnamefont {M.}~\bibnamefont {Di~Ventra}}, \bibinfo
 {author} {\bibfnamefont {M.}~\bibnamefont {Sanz}}, \ and\ \bibinfo {author}
 {\bibfnamefont {E.}~\bibnamefont {Solano}},\ }\href@noop {} {\bibfield
 {journal} {\bibinfo {journal} {Sci. Rep.}\ }\textbf {\bibinfo {volume}
 {6}},\ \bibinfo {pages} {29507} (\bibinfo {year} {2016})}\BibitemShut
 {NoStop}%
\bibitem [{\citenamefont {Kaplan}\ \emph {et~al.}(1976)\citenamefont {Kaplan},
 \citenamefont {Chi}, \citenamefont {Langenberg}, \citenamefont {Chang},
 \citenamefont {Jafarey},\ and\ \citenamefont {Scalapino}}]{Kap76}%
 \BibitemOpen
 \bibfield {author} {\bibinfo {author} {\bibfnamefont {S.~B.}\ \bibnamefont
 {Kaplan}}, \bibinfo {author} {\bibfnamefont {C.~C.}\ \bibnamefont {Chi}},
 \bibinfo {author} {\bibfnamefont {D.~N.}\ \bibnamefont {Langenberg}},
 \bibinfo {author} {\bibfnamefont {J.~J.}\ \bibnamefont {Chang}}, \bibinfo
 {author} {\bibfnamefont {S.}~\bibnamefont {Jafarey}}, \ and\ \bibinfo
 {author} {\bibfnamefont {D.~J.}\ \bibnamefont {Scalapino}},\ }\href {\doibase
 10.1103/PhysRevB.14.4854} {\bibfield {journal} {\bibinfo {journal} {Phys.
 Rev. B}\ }\textbf {\bibinfo {volume} {14}},\ \bibinfo {pages} {4854}
 (\bibinfo {year} {1976})}\BibitemShut {NoStop}%
\bibitem [{\citenamefont {Barends}\ \emph {et~al.}(2008)\citenamefont
 {Barends}, \citenamefont {Baselmans}, \citenamefont {Yates}, \citenamefont
 {Gao}, \citenamefont {Hovenier},\ and\ \citenamefont {Klapwijk}}]{Bar08}%
 \BibitemOpen
 \bibfield {author} {\bibinfo {author} {\bibfnamefont {R.}~\bibnamefont
 {Barends}}, \bibinfo {author} {\bibfnamefont {J.~J.~A.}\ \bibnamefont
 {Baselmans}}, \bibinfo {author} {\bibfnamefont {S.~J.~C.}\ \bibnamefont
 {Yates}}, \bibinfo {author} {\bibfnamefont {J.~R.}\ \bibnamefont {Gao}},
 \bibinfo {author} {\bibfnamefont {J.~N.}\ \bibnamefont {Hovenier}}, \ and\
 \bibinfo {author} {\bibfnamefont {T.~M.}\ \bibnamefont {Klapwijk}},\ }\href
 {\doibase 10.1103/PhysRevLett.100.257002} {\bibfield {journal} {\bibinfo
 {journal} {Phys. Rev. Lett.}\ }\textbf {\bibinfo {volume} {100}},\ \bibinfo
 {pages} {257002} (\bibinfo {year} {2008})}\BibitemShut {NoStop}%
 \bibitem [{\citenamefont {Kusunose}(2004)}]{Kus04}%
 \BibitemOpen
 \bibfield {author} {\bibinfo {author} {\bibfnamefont {H.}~\bibnamefont
 {Kusunose}},\ }\href {\doibase 10.1103/PhysRevB.70.054509} {\bibfield
 {journal} {\bibinfo {journal} {Phys. Rev. B}\ }\textbf {\bibinfo {volume}
 {70}},\ \bibinfo {pages} {054509} (\bibinfo {year} {2004})}\BibitemShut
 {NoStop}%
 \bibitem [{\citenamefont {Rabani}\ \emph {et~al.}(2008)\citenamefont {Rabani},
 \citenamefont {Taddei}, \citenamefont {Bourgeois}, \citenamefont {Fazio},\
 and\ \citenamefont {Giazotto}}]{Rab08}%
 \BibitemOpen
 \bibfield {author} {\bibinfo {author} {\bibfnamefont {H.}~\bibnamefont
 {Rabani}}, \bibinfo {author} {\bibfnamefont {F.}~\bibnamefont {Taddei}},
 \bibinfo {author} {\bibfnamefont {O.}~\bibnamefont {Bourgeois}}, \bibinfo
 {author} {\bibfnamefont {R.}~\bibnamefont {Fazio}}, \ and\ \bibinfo {author}
 {\bibfnamefont {F.}~\bibnamefont {Giazotto}},\ }\href {\doibase
 10.1103/PhysRevB.78.012503} {\bibfield {journal} {\bibinfo {journal} {Phys.
 Rev. B}\ }\textbf {\bibinfo {volume} {78}},\ \bibinfo {pages} {012503}
 (\bibinfo {year} {2008})}\BibitemShut {NoStop}%
 \bibitem [{\citenamefont {Solinas}\ \emph {et~al.}(2016)\citenamefont
 {Solinas}, \citenamefont {Bosisio},\ and\ \citenamefont {Giazotto}}]{Sol16}%
 \BibitemOpen
 \bibfield {author} {\bibinfo {author} {\bibfnamefont {P.}~\bibnamefont
 {Solinas}}, \bibinfo {author} {\bibfnamefont {R.}~\bibnamefont {Bosisio}}, \
 and\ \bibinfo {author} {\bibfnamefont {F.}~\bibnamefont {Giazotto}},\ }\href
 {\doibase 10.1103/PhysRevB.93.224521} {\bibfield {journal} {\bibinfo
 {journal} {Phys. Rev. B}\ }\textbf {\bibinfo {volume} {93}},\ \bibinfo
 {pages} {224521} (\bibinfo {year} {2016})}\BibitemShut {NoStop}%


\end{thebibliography}

%

\end{document}